\documentclass[a4paper,11pt]{article}
\pdfoutput=1
\usepackage{jcappub}

\usepackage{tikz,xcolor,hyperref}
\usepackage{amsmath, amssymb, amsthm, graphicx, epsfig, fancyhdr,epsfig, slashed}
\usepackage[normalem]{ulem}
\usepackage{tikzsymbols}
\usepackage{natbib}
\usepackage{float}
\usepackage{bm}
\usepackage{adjustbox}

\usepackage[bitstream-charter]{mathdesign}
\usepackage[nointegrals]{wasysym} 

\DeclareSymbolFont{myletters}{OML}{ztmcm}{m}{it}
\DeclareMathSymbol{\uplambda}{\mathord}{myletters}{"15}
\DeclareMathSymbol{\upxi}{\mathord}{myletters}{"18}

\usepackage{amsfonts}

\renewcommand{\figurename}{FIGURE}

\usepackage{lmodern} 
\usepackage{amsmath}                                                    
\usepackage{amsthm}                                                     
\usepackage{amssymb}
\usepackage{multirow}
\usepackage[nameinlink]{cleveref}
\Crefname{figure}{Fig.}{Figs.}

\usepackage{adjustbox}
\usepackage{color, colortbl} 
\definecolor{Gray}{gray}{0.9}
\usepackage{ragged2e}
\usepackage{titlesec}
\usepackage[includemp,
paperwidth=21.5cm,
paperheight=29.7cm,
top=2.30cm,
bottom=2.3cm,
inner=3cm,
outer=2.0cm,
marginparwidth=0.3cm,
marginparsep=0.3cm]{geometry}
\newcolumntype{P}[1]{>{\centering\arraybackslash}p{#1}}
\newcolumntype{M}[1]{>{\centering\arraybackslash}m{#1}}
\usepackage{enumitem}
\usepackage{cellspace}
\usepackage{makecell}
\usepackage{setspace}
\usepackage[nameinlink]{cleveref}
\Crefname{figure}{Fig.}{Figs.}

\usepackage{subfig}

\def\beq{\beq\begin{align}}
\def\eeq{\end{align}\eeq}

\def\beq{\begin{equation}\begin{align}}
\def\eeq{\end{align}\end{equation}}

\begin{document}
\title{Primordial Black Holes and Scalar-induced Gravitational Waves in Sneutrino Hybrid Inflation}

\author[]{Adeela Afzal $^{a\dagger}$,}
\author[]{Anish Ghoshal $^{b\ddagger}$,}
\author[]{Stephen F. King $^{c*}$}

\affiliation[a]{\it Department of Physics,  Quaid-i-Azam University, Islamabad, 45320, Pakistan}
\affiliation[b]{\small\it Institute of Theoretical Physics, Faculty of Physics, University of Warsaw, ul. \\ Pasteura 5, 02-093 Warsaw, Poland}
\affiliation[c]{\small\it School of Physics and Astronomy, University of Southampton, Southampton \\ SO17 1BJ, United Kingdom}

\emailAdd{$^\dagger$adeelaafzal555@gmail.com}
\emailAdd{$^\ddagger$anish.ghoshal@fuw.edu.pl}
\emailAdd{$^*$king@soton.ac.uk}

\abstract{We investigate the possibility that primordial black holes (PBHs) can be formed from large curvature perturbations generated during the waterfall phase transition in a supersymmetric scenario where sneutrino is the inflaton in a hybrid inflationary framework. We obtain a spectral index ($n_s \simeq 0.966$), and a tensor-to-scalar ratio ($r\simeq 0.0056-10^{-11}$), consistent with the current Planck data satisfying PBH as dark matter (DM) and detectable Gravitational Wave (GW) signal. Our findings show that the mass of PBH and the peak in the GW spectrum is correlated with the right-handed (s)neutrino mass. We identify parameter space where PBHs can be the entire DM candidate of the universe (with mass $10^{-13}\, M_\odot$) or a fraction of it. This can be tested in future observatories, for example, with amplitude $\Omega_{\rm GW}h^2$ $\sim 10^{-9}$ and peak frequency $f\sim 0.1$ Hz in LISA and $\Omega_{\rm GW}h^2 \sim 10^{-11}$ and peak frequency of $\sim 10$ Hz in ET via second-order GW signals.
We study two models of sneutrino inflation: Model$-1$ involves canonical sneutrino kinetic term which predicts the sub-Planckian mass parameter $M$, while the coupling between a gauge singlet and the waterfall field, $\beta$, needs to be quite large whereas, for the model$-2$ involving $\alpha-$attractor canonical sneutrino kinetic term, $\beta$ can take a natural value. Estimating explicitly, we show that both models have mild fine-tuning. We also derive an analytical expression for the power spectrum in terms of the microphysics parameters of the model like (s)neutrino mass, etc. that fits well with the numerical results. The typical reheat temperature for both the models is around $10^{7}-10^{8}$~GeV suitable for non-thermal leptogenesis.}

\maketitle

\section{Introduction}
\label{sec:intro}


Following recent remarkable data from cosmic microwave background radiation (CMBR) acquired by the Planck satellite, cosmic inflation has become the paradigm for early universe cosmology. Moreover, due to the vacuum energy scale during inflation being approximately $10^{14}$ GeV (largest energy scale allowed by CMB) is close to the expected Grand Unification scale and seesaw scale, such measurements are a pathway to directly probe into ultra-violate particle physics.

Alongside the CMB scale and the physics during inflation, there has also been a great growing interest in the late stages of the inflationary paradigm, particularly the investigating scenarios where considerably large peaks in the amplitude of perturbations could be realized. This is because such peaks could lead to the formation of primordial black holes (PBHs) which has gained a lot of interest after the observations of the supermassive ~\cite{LyndenBell:1969yx, Kormendy:1995er} and stellar-mass black hole (BH) merger events detected via observing gravitational wave (GW) in LIGO-VIRGO-KAGRA~\cite{LIGOScientific:2016dsl, LIGOScientific:2021djp}. Such a large primordial density fluctuation could generate a stochastic gravitational wave background (SGWB) when PBHs are generated via the collapse of high-dense regions~\cite{Hawking:1971ei,Carr:1974nx,Carr:1975qj}. Typically, this requires the inflaton to go through a very flat potential during inflation (see ref.~\cite{Escriva:2022duf} for a recent review on PBHs) and involves some higher degrees of fine-tuning; see ref.\cite{Cole:2023wyx} for details. This was first studied decades ago in ref.~\cite{Garcia-Bellido:1996mdl} in the context of a hybrid inflation scenario \cite{Linde:1991km,Linde:1993cn}.
  It was recently shown that hybrid inflation naturally reduces the fine-tuning involved in producing PBH as dark matter (DM) \cite{Afzal:2024xci,Spanos:2021hpk}\footnote{Some other studies on PBHs include: Supermassive primordial black holes in the multiverse: for nano-Hertz gravitational wave and high-redshift JWST galaxies has been studied in~\cite{Huang:2023chx}. The study of induced gravitational waves from flipped SU(5) superstring theory at nHz has been explored in \cite{Basilakos:2023jvp}.}.

In this paper, we reconsider the possibility that the inflaton sector is responsible for generating the tiny neutrino masses required to explain several neutrino oscillation experiments. Heavy singlet neutrinos with scalar partners called sneutrinos with $10^{10} - 10^{15}$ GeV, lie in the range where the inflaton mass may lie.  In this paper, we discuss two supersymmetric scenarios where the lightest heavy singlet (s)neutrino drives inflation. As we will see, this scenario constrains in interesting ways many of the $18$ parameters of the minimal seesaw model for generating three non-zero light neutrino masses. This minimal (s)neutrino inflationary scenario can be rescued within the framework of hybrid inflation along with a successful neutrino oscillation data explanation. In terms of the early universe, this scenario provides a natural way to accommodate baryogenesis via leptogenesis due to sneutrino decay \cite{Ellis:2005uk}. We will show that the mass of PBH and the peak in the GW spectrum are correlated with the right-handed (s)neutrino mass given that (s)neutrino is responsible for driving the cosmic inflaton phase.

As previously demonstrated in the literature, the waterfall transition happening in the last stages of inflation, can have a flat potential to generate large curvature perturbations at small scales~\cite{Clesse:2010iz,Kodama:2011vs,Mulryne:2011ni}. These large density perturbations, lead to a larger amplitude of curvature perturbations and eventually collapses into a PBH. As demonstrated in refs.~\cite{Clesse:2015wea,Kawasaki:2015ppx}, this usually leads to the PBH overproduction of astrophysical size. However, this can be circumvented and the PBH abundance modulated with observable Scalar Induced Gravitational Wave (SIGW) signals using a slightly modified waterfall field potential \cite{Braglia:2022phb}. The PBH overproduction issue is addressed by introducing the linear term along with consistent CMB predictions. We present in our analysis the analytical expressions for the power spectrum (see \cref{PBHOverpropsana}) in terms of the hybrid inflation microphysics parameters taking care of the PBH overproduction issue. Since the crucial process of PBH generation estimates is subject to multiple significant uncertainties, we study it in less detail. However, we study in detail the characterization of the SGWB induced at the second order by the large curvature perturbations (at small length scales) during horizon reentry in the
radiation-dominated era\footnote{See ref. \cite{Moursy:2024hll,Cui:2021are} for other ways to generate PBH during hybrid inflation scenarios. For grand unification in hybrid inflation, we refer the reader to~\cite{Buchmuller:2014dda}. A detailed analysis of the T and E models of hybrid inflation is discussed in~\cite{Pallis:2022mzh}.}.  

In a nutshell, we show that supersymmetric hybrid inflation involving neutrino mass could be tested via PBH and GW observations at energy scales beyond the TeV limit of collider physics. The flow of the paper is as follows: in \cref{PBHOverpro} we presented a toy model of hybrid inflation and the numerical framework for the power spectrum is given in \cref{sec:numpowerspec}. The analytical calculations for the power spectrum are given in \cref{PBHOverpropsana}. We analyze the toy model in a realistic supersymmetric framework in \cref{realsiticframe}. The PBH abundance and SIGWs are calculated in \cref{sec:PBHabund} and \cref{sec:SIGW} respectively. We modify our model in terms of $\alpha$-attractor framework in \cref{sec:alphaattrac} and present the fine-tuning estimate for both cases in \cref{sec:finetune}. We conclude in \cref{sec:concl}.
\section{ {\it Model$-1$}: Toy Model Hybrid Inflation}
\label{PBHOverpro}

\begin{figure}[tbh!]
    \centering
    \includegraphics[width=0.9\linewidth]{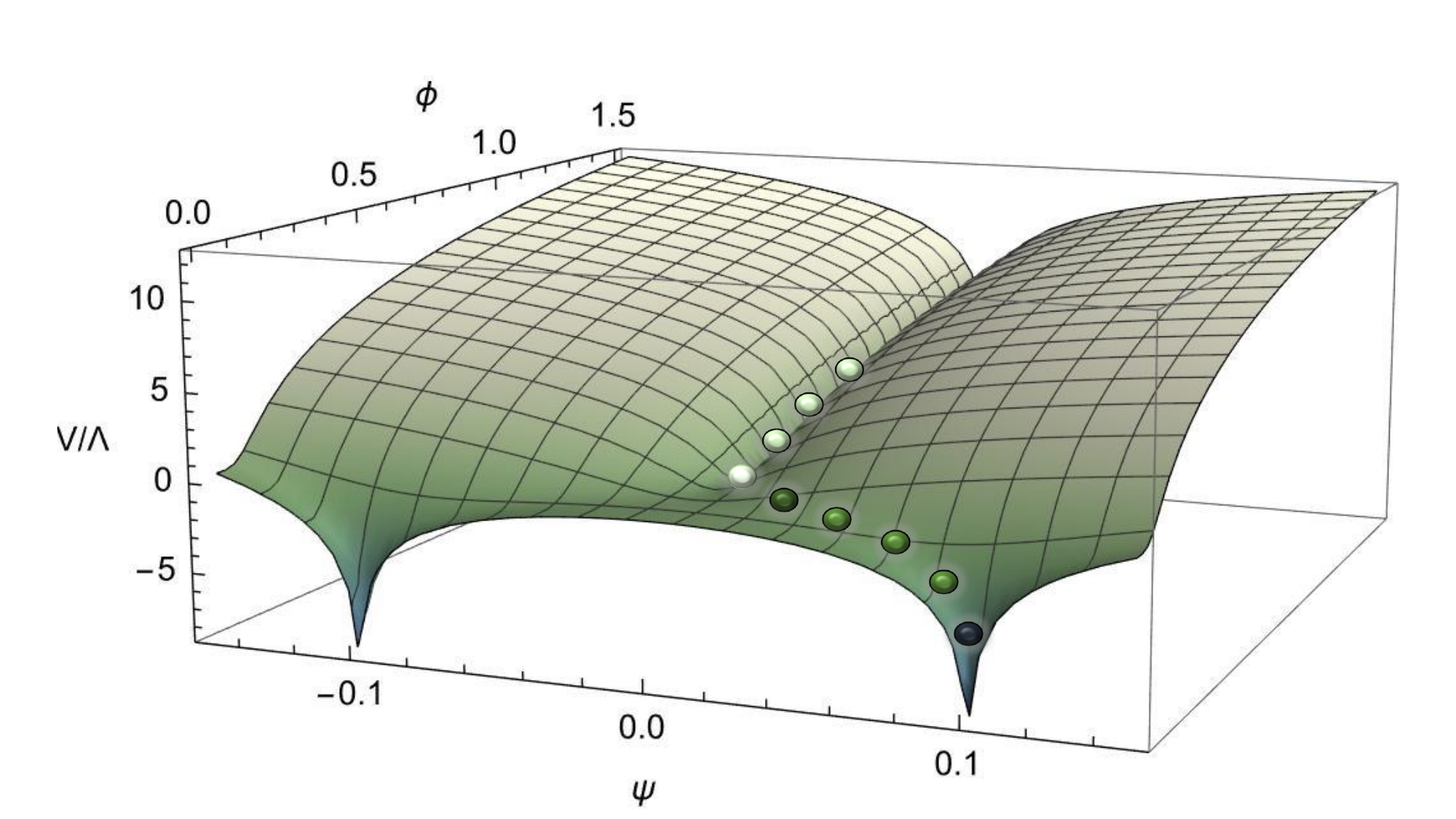}
    \caption{\it A schematic picture of hybrid inflation potential given in \cref{potentialCl}. The light green bullets show the inflationary trajectory of the inflaton $\phi$, the white bullet is the critical point and the dark green bullets show the waterfall regime.}
    \label{fig:potential}
\end{figure}

Our focus in this paper is to study the footprints of supersymmetry in the SIGWs and PBHs. For this purpose, let us consider a toy model with hybrid inflationary potential. The critical point of instability, where the two scalar fields in the hybrid inflation become unstable, plays a crucial role in PBH production. The hybrid inflationary potential close to the critical point of instability can be written as 
\begin{align}
\label{potentialCl}
    V=\Lambda\left(\left(1-\dfrac{\psi^2}{M^2}\right)^2+\dfrac{\phi-\phi_c}{m_1}- \dfrac{\left(\phi-\phi_c\right)^2}{m_2^2}+\dfrac{\phi^4\psi^2}{M^2\,\phi_c^4}+\dfrac{\psi}{b}\right).
\end{align}
Here, $\phi$ is the inflaton field, $\psi$ is the waterfall field, $\phi_c$ is the critical point below which the potential develops a tachyonic instability, forcing
the field trajectories to reach one of the global minima,
located at $\phi=0,\,\, \psi=\pm M$, $m_1$, $m_2$ are the dimensionful mass parameters and $\Lambda$ is the non-zero vacuum energy\footnote{The minimum of the whole potential is $V_{\rm min} = \pm\Lambda(M/b)$.
Now, $\Lambda$ is the inflation energy density scale because $H^2=\Lambda/3$ as mentioned before \cref{ph1kg}, however, $b$ is not too small (in Table 1 and Fig. 3 we will see that $b \sim 10^9 m_\text{Pl}$ and $M=0.1\, m_\text{Pl}$ (Table 1). This means that
$V_\text{min} \sim \pm 10^{-22} m_\text{Pl}^4$. Although this is huge compared to the vacuum density of $\Lambda$CDM  we can always adjust the unknown cosmological constant to cancel the vacuum density of the deeper minimum.}. Hybrid inflation models, where PBH formation was first presented rigorously \cite{Clesse:2015wea}, predict the overabundance of DM. 
We have added a linear term in the waterfall field that does not allow 
 $\psi$ to relax exactly at zero but slightly displace depending
upon the sign. of the co-efficient of the linear term $1/b$. 
In this way, one can control the peak of the power spectrum and avoid PBH overproduction. 
A schematic picture of the model is shown in \cref{fig:potential}.

The slow-roll parameters are given by, \cite{Chatterjee:2017hru},
\begin{align}
\label{eq:slowroll}
    \epsilon_V = \dfrac{m_\text{Pl}^2}{2}\left(\dfrac{\partial_{X} V}{V}\right)^2,\,\,\,\,\,\eta_V = m_\text{Pl}^2\left(\dfrac{\partial_{X}^2 V}{V}\right).
\end{align}

\begin{figure}[tbh!]
    \centering
    \includegraphics[width=0.7\linewidth]{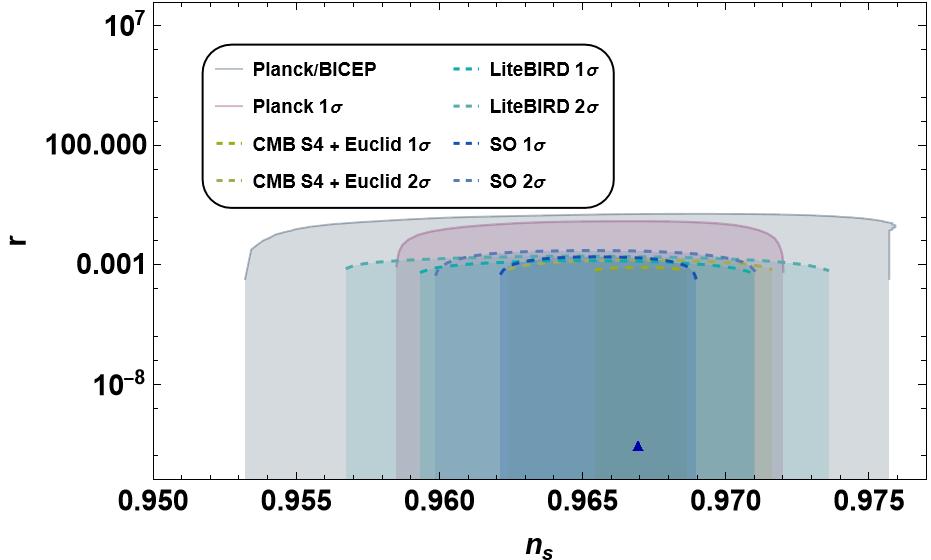}
    \caption{\it \it Tensor-to-scalar ratio $r$ vs. scalar spectral index $n_s$ for the corresponding parameter sets given in \cref{parmsetsCl}. The solid contours are the current Planck bounds \cite{Planck:2018jri}, Planck/BICEP \cite{BICEP:2021xfz,Planck:2018vyg,BICEPKeck:2022mhb} and the dashed shaded region indicates the future proposed experiments (LiteBIRD, CMB-Euclid, Simons Observatory (SO)) \cite{laureijs2011euclid, SimonsObservatory:2018koc, LiteBIRD:2020khw}.}
    \label{fig:rnsClesse}
\end{figure}

Here, $m_\text{Pl}\simeq 2.43 \times 10^{18}$ 
 GeV is the reduced Planck mass and $\partial_X \equiv  \partial/\partial X$ with $X=\{\phi,\psi\}$ is the field derivative. In the slow roll limit, the spectral index $n_s$ is given by \cite{Chatterjee:2017hru},
\begin{align}
\label{nsact}
    n_s &= 1-6\,\epsilon_V + 2\,\eta_V.
\end{align}
The central measurements by Planck 2018 in the $\Lambda$CDM model are; $n_s = 0.9647 \pm 0.012$ and the tensor to scalar ratio $r=16\,\epsilon_V < 0.035$ 
at $95\%$ C.L \cite{Planck:2018vyg}. All the values are measured at the pivot scale, $k_0=0.05\,\text{Mpc}^{-1}$. The subscript $0$ from here and onward indicates the value at the pivot scale. 
Using \cref{eq:slowroll}, along the valley when $X=\phi$ one obtains the slow roll parameters to be, 

\begin{align}
\label{epetCL}
	\left.\epsilon_V\right|_{\phi=\phi_{c}}\simeq \dfrac{m_\text{Pl}^2}{2\,m_1^2}\,\,\,\,\,\,\,\,\,\text{and}\,\,\,\,\,\,\,\, \left.\eta_V\right|_{\phi=\phi_{c}}\simeq-\dfrac{2\,m_\text{Pl}^2}{m_2^2}.
\end{align}
The amplitude of the scalar power spectrum along the valley is given by,
\begin{align}
\label{pivot}
	A_s(k_0)=\left.\dfrac{V}{24\,\pi^2m_\text{Pl}^4\,\epsilon_V}\right|_{\phi=\phi_{0}}.
\end{align}
This amplitude is fixed by Planck's result $A_s(k_0)=2.198\times10^{-9}$.  Considering the vacuum energy to be the dominant source at the pivot scale, substituting the $\epsilon_V$ from \cref{epetCL} and the amplitude of the scalar power spectrum in \cref{pivot}, we obtained the relation between $\Lambda$ and $m_1$, 
\begin{align}
\label{lambda}
\Lambda\simeq2.198\times10^{-9}\times 12\,\pi^2\left(\dfrac{m_\text{Pl}^6}{m_1^2}\right).
\end{align} 
In our region of interest $m_1 \gg m_2$, following \cref{nsact}, the scalar spectral index is now written as
\begin{align}
\label{CLns}
	n_s
	\simeq1-\dfrac{4\,m_\text{Pl}^2}{m_2^2}.
\end{align}
The value of $n_s$ in the $\Lambda$CDM model is $n_s = 0.9665 \pm 0.0038$ \cite{Planck:2018jri} that fixes $m_2\simeq 11\,m_\text{Pl}$. The corresponding value of $n_s$ is shown in \cref{fig:rnsClesse} for the benchmark point (BP) in \cref{parmsetsCl} along with present and future plan experiments. The relevant number of e-folds, $N_0$, before the end of inflation are,
\begin{align}\label{Ngen}
	N_0 \simeq \left( \frac{1}{m_\text{Pl}}\right) ^{2}\int_{X_e}^{X_{0}}\left( \frac{V}{%
		V_X}\right) dX,
\end{align}
where $X_e$ is the field value at the end of inflation which is fixed by the breakdown of the slow-roll approximation. The parameter set for different model variables we consider here is given in \cref{parmsetsCl}.
\begin{table}[tbh!]
    \fontsize{12pt}{20pt}
    \caption{\it Benchmark points for model parameters \cref{potentialCl}} 
    \centering 
    \begin{adjustbox}{max width=\columnwidth}
        \begin{tabular}{|M{1.3cm} |M{1.4cm} |M{1.3cm}|M{2.0cm} |M{1.5cm} |M{2.2cm} |M{2.4cm}|M{1.1cm}|M{1.1cm}|M{2.1cm}|}
            \hline
            \bf{Model}& \bf{$M/m_\text{Pl}$}&\bf{$\phi_c/m_\text{Pl}$}&\bf{$m_1/m_\text{Pl}$}&\bf{$m_2/m_\text{Pl}$}&\bf{$b/m_\text{Pl}$}&\bf{$\phi_i$}&\bf{$\psi_i$}&\bf{$N_k$}&\bf{$r$}\\ [0.6ex] 
            \hline\hline
            $\text{BP}$ &$0.1$ &$0.1$ &$3.00\times 10^{5}$ &$11$  & $-8.00\times 10^{9}$ & $ \phi_c (1 + 0.0011) $ &$\psi_0$&$55$&$2.89\times 10^{-11}$\\
            \hline
        \end{tabular}
    \end{adjustbox}
    \label{parmsetsCl}
\end{table}

\medskip

\section{Numerical Treatment of Scalar Perturbations}
\label{sec:numpowerspec}
The Klein-Gordon classical background equations of motion in the number of e-fold times are given by \cite{Clesse:2013jra},
\begin{align}
\label{bac}	\phi^{''}+\left(\dfrac{H^{'}}{H}+3\right)\phi^{'}+\dfrac{V_{\phi}}{H^2}=0,\,\,\,\,\,\,\,\,\,\,\,\,\,\,
	\psi^{''}+\left(\dfrac{H^{'}}{H}+3\right)\psi^{'}+\dfrac{V_{\psi}}{H^2}=0.
\end{align}
Here, $V_X=dV/dX$ where, $X=\{\phi,\psi\}$, prime is the derivative with respect to the number of e-folds and the Hubble rate $H$ is defined as $H^2=2\,V/(6-\phi^{'2}-\psi^{'2})$. The evolution of the field from pivot scale till the end of inflation, $\epsilon_V=1$, is shown in \cref{fig:phipsi}. The scalar perturbations of the FLRW metric in longitudinal gauge can be expressed as \cite{Clesse:2013jra},
\begin{align}
	ds^2=a^2\left[(1+2\,\Phi_{\text{B}})d\tau^2+\left[(1-2\,\Psi_{\text{B}})\delta_{ij}+\dfrac{h_{ij}}{2}\right]dx^idx^j\right],
\end{align}
where $\tau$ is the conformal time related to cosmic time via scale factor $a$, $dt=a\,d\tau$, $\Phi_{\text{B}}$ and $\Psi_{\text{B}}$ are the Bardeen potentials and $h_{ij}$ is the transverse-traceless tensor metric perturbation. We work in a conformal Newtonian gauge such that $\Phi_{\text{B}}=\Psi_{\text{B}}$. The scalar perturbations are defined as \cite{Ringeval:2007am, Clesse:2013jra},
\begin{align}
	\delta X_i^{''}+(3-\epsilon)\delta X_i^{'}+\sum_{j=1}^{2}\dfrac{1}{H^2}V_{X_i X_j}\delta X_j+\dfrac{k^2}{a^2H^2}\delta X_i=4\Phi^{'}_{\text{B}}X^{'}_i-\dfrac{2\,\Phi_{\text{B}}}{H^2}V_{X_i}.
\end{align}
Here $X$ with the subscript $(i, j)$ refers to the fields ($\phi$, $\psi$), $k$ is the comoving wave number, the equation of motion for $\Phi_{\text{B}}$ is given by,
\begin{align}
	\Phi^{''}_{\text{B}}+(7-\epsilon)\Phi^{'}_{\text{B}}+\left(\dfrac{2V}{H^2}+\dfrac{k^2}{a^2H^2}\right)\Phi_{\text{B}}+\dfrac{V_{X_i}}{H^2}\delta X_i=0.
\end{align} 

The initial conditions (i.c) for field perturbations in e-fold time are given as,
\begin{align}
	\delta X_{i,\text{i.c}}=\dfrac{1}{a_{\text{i.c}}\sqrt{2 k}},\,\,\,\,\,\,\,\,\,\,\,\,\,\,
	\delta X_{i,\text{i.c}}^{'}=-\dfrac{1}{a_{\text{i.c}}\sqrt{2 k}} \left(1+\iota\dfrac{k}{a_{\text{i.c}}H_{\text{i.c}}}\right).
\end{align}
The initial conditions for the Bardeen potential and its derivative are given by,
\begin{align}
	\Phi_{\text{B,i.c}}&=\sum_{j=1}^{2}\dfrac{\left(H_{\text{i.c}}^2 X_{i,\text{i.c}}^{'}\delta X_{i,\text{i.c}}^{'}+\left(3H_{\text{i.c}}^2 X_{i,i.c}^{'}+V_{X_i,\text{i.c}}\right)\delta X_{i,\text{i.c}}\right)}{2H_{\text{i.c}}^2\left(\epsilon_{\text{i.c}}-\dfrac{k^2}{a_{\text{i.c}}^2H^2_{\text{i.c}}}\right)},\\ \notag
	\Phi^{'}_{\text{B,i.c}}&=-\Phi_{\text{B,i.c}}+\sum_{j=1}^{2}\dfrac{X_{i,\text{i.c}}^{'}\delta X_{i,\text{i.c}}}{2}.
\end{align}
The scalar power spectrum in terms of curvature perturbations $\zeta$ is given by,
\begin{align}
\label{Powspec}
	P_R(k)=\dfrac{k^3}{2\pi^2}\left|\zeta\right|^2=\dfrac{k^3}{2\pi^2}\left|\Phi_{\text{B}}+\dfrac{\sum_{i=1}^{2}X_i^{'}\delta X_i}{\sum_{j=1}^{2}X^{'2}_j}\right|^2.
\end{align} 
Later we used this relation to numerically evaluate the power spectrum.

\medskip

 \section{Power Spectrum Analytical Expression}
 \label{PBHOverpropsana}

\begin{figure}[t]
    \centering
    \includegraphics[width=0.49\linewidth,height=5cm]{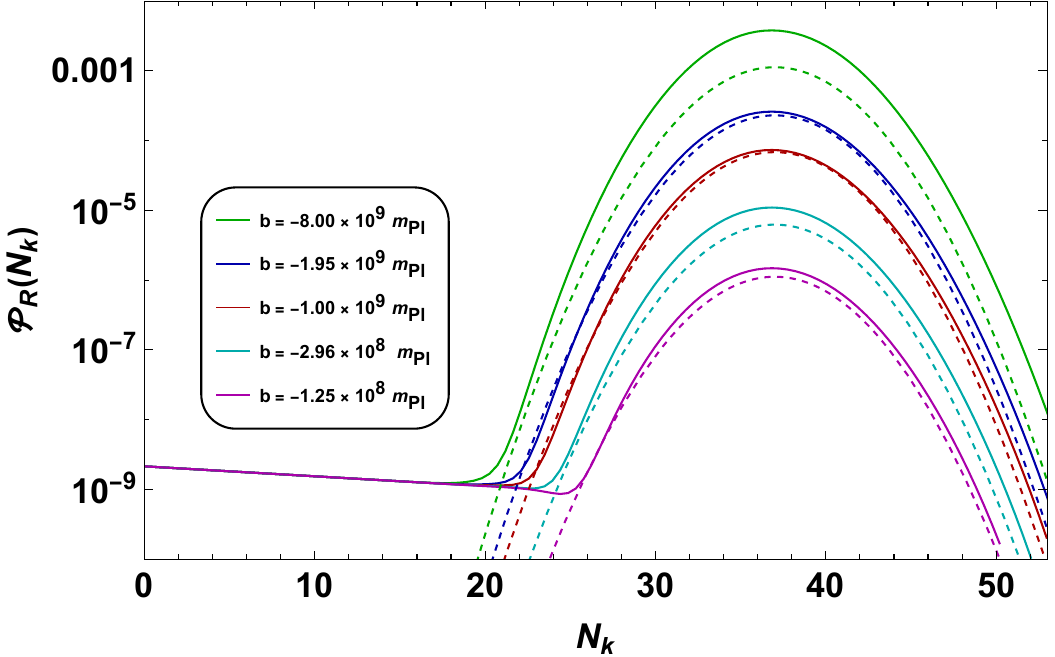}
    \includegraphics[width=0.49\linewidth,height=5cm]{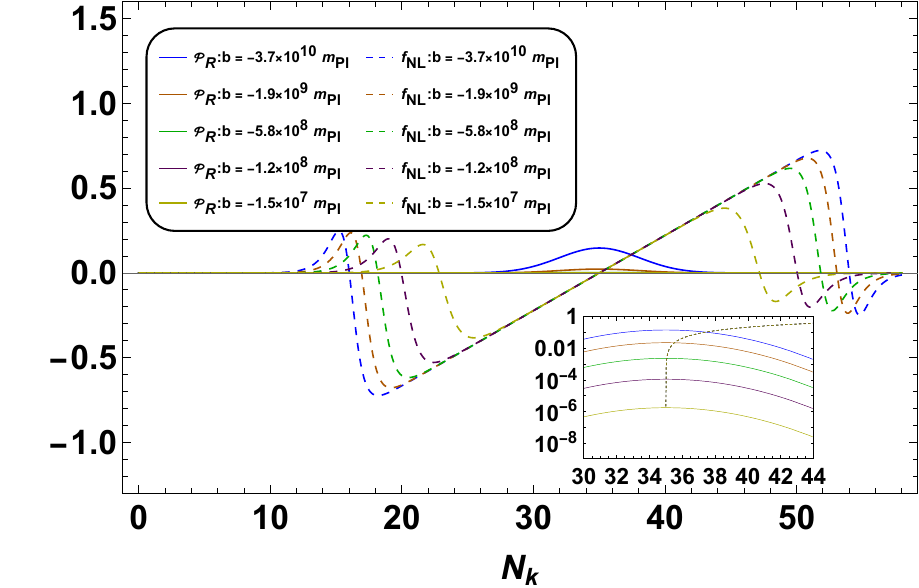}
    \caption{\it Left panel presents the power spectrum by solving the exact linear scalar perturbation equations (solid) \cref{Powspec} and for the analytical expression (dashed) \cref{PSanalyt} for the BP in \cref{parmsetsCl}. The $f_\text{NL}$ parameter given in \cref{fNL} along with the power spectrum is presented in the right panel for the BP given in \cref{parmsetsCl} along with the variation of the coefficient of the linear term. The inset plot shows the peak value of the power spectrum on the log axis.}
\label{fig:AnaExactPS_k_Fudge_factor}
\end{figure}

 In this section, we provide an analytical expression for the scalar power spectrum following Ref. \cite{Clesse:2015wea}.
To calculate the analytical expression for the power spectrum, let us begin with the slow roll approximation (in units $m_\text{Pl}=1$) under which the Klein-Gordon equations with respect to the number of e-folds can be written as,
\begin{align}
    3\,H^2\phi^{'}& =-V_\phi=-\Lambda \left(\dfrac{1}{m_1}+\dfrac{8\,\phi^3\psi^2}{M^2\,\phi_c^4}\right),\\ \notag3\,H^2\psi^{'}&=-V_\psi=-\Lambda\left(b^{-1}+\dfrac{4\,\phi^4\psi}{M^2\,\phi_c^4}-\dfrac{4\,\psi}{M^2}\left(1-\dfrac{\psi^2}{M^2}\right)\right).
\end{align}
There are three different phases in the waterfall regime and only phase-1 is important \cite{Clesse:2015wea} during which, the above equations reduce to,
\begin{align}
\label{ph1kgeqn}
    3\,H^2\phi^{'}& =-\Lambda \left(\dfrac{1}{m_1}\right),\\ \notag3\,H^2\psi^{'}&=-\Lambda\left(b^{-1}+\dfrac{4\,\phi^4\psi}{M^2\,\phi_c^4}-\dfrac{4\,\psi}{M^2}\right).
\end{align}
Defining $H^2=\Lambda/3$, \cref{ph1kgeqn} becomes,
\begin{align}
\label{ph1kg}
    \phi^{'}& =-\left(\dfrac{1}{m_1}\right),\\ \notag
    \psi^{'}&=-\left(b^{-1}+\dfrac{4\,\phi^4\psi}{M^2\,\phi_c^4}-\dfrac{4\,\psi}{M^2}\right).
\end{align}
Let us assume the solution,
\begin{align}
\label{solkg1}
    \phi=\phi_c\,e^{\xi},\,\,\,\,\,\,\,\,\,\,\,\,\,\,\,\,\,\,
    \psi=\psi_0\,e^{\chi}.
\end{align}
Under the slow roll approximation, during the waterfall, $|\xi|\ll1$, one can write;
\begin{align}
\label{solkg}
\phi\simeq\phi_c\,\left(1+\xi\right),\,\,\,\,\,\,\,\,\,\,\,\,\,\,\,\,\,\,
    \psi=\psi_0\,e^{\chi}.
\end{align}
Here, $\psi_0$ is the auxiliary field distribution width at the critical point of instability given by \cite{Garcia-Bellido:1996mdl},
\begin{align}
   \label{auxfield} \psi_0=\sqrt{\dfrac{\Lambda\,M\,\sqrt{2\,\phi_c\,m_1}}{92\,\pi^{3/2}\,m_\text{Pl}^4}}.
\end{align}
Under solutions \cref{solkg} and taking into account $|\xi|\ll1$, one can write \cref{ph1kg} as,
\begin{align}
\label{ph1kgsol}
    \xi^{'}& \simeq -\left(\dfrac{1}{m_1\,\phi_c}\right),\\\notag
    \chi^{'}&=-\left(\dfrac{b^{-1}}{\psi_0\,e^\chi}+\dfrac{4\,\phi_c^4}{M^2\,\phi_c^4}\,\left(1+\xi\right)^4-\dfrac{4}{M^2}\right)\\\notag
    &\simeq-\left(\dfrac{b^{-1}}{\psi_0\,e^\chi}+\dfrac{4\,\phi_c^4}{M^2\,\phi_c^4}\,\left(1+4\,\xi\right)-\dfrac{4}{M^2}\right).
\end{align}
 This gives,
\begin{align}
\label{xichi}
    \dfrac{d\xi}{d\chi}=\dfrac{1}{m_1\,\phi_c\left(\dfrac{b^{-1}}{\psi_0\,e^\chi}+\dfrac{4\,\phi_c^4}{M^2\,\phi_c^4}\,\left(1+4\,\xi\right)-\dfrac{4}{M^2}\right)}.
\end{align}
Solving \cref{ph1kgsol} for $\xi_N$ one gets,
\begin{align}
\label{xiefolds}
    \xi\simeq -\dfrac{1}{m_1\,\phi_c}\, N.
\end{align}
Or one can write,
\begin{align}
\label{efoldsph1}
    N\simeq-m_1\,\phi_c\,\xi.
\end{align}
The above equation gives the number of e-folds in phase-1. Consider \cref{xichi}, it is not easy to solve analytically. Let us consider the approximation $\psi_0\,e^{\chi}\simeq\psi_0$, assuming $\chi\ll\xi$ , 
\begin{align}
\label{xichi1}
    \dfrac{d\xi}{d\chi}\simeq\dfrac{1}{m_1\,\phi_c\left(\dfrac{b^{-1}}{\psi_0}+\dfrac{4}{M^2}\,\left(4\,\xi\right)\right)}.
\end{align}
This implies,
\begin{align}
    \xi\simeq \dfrac{-b^{-1}\,M^4\,m_1+M^2\sqrt{m_1\,\left(b^{-2}\,M^4\,m_1+32\,\phi_c\,\psi_0^2\,\chi\right)}}{16\,\psi_0\,m_1\phi_c^2}.
\end{align}
Therefore, the number of e-folds in phase-1 are,
\begin{align}
\label{efoldsph1}
    N_1\simeq -\xi\,m_1\,\phi_c\simeq\dfrac{-b^{-1}\,M^4\,m_1+M^2\sqrt{m_1\,\left(b^{-2}\,M^4\,m_1+32\,\phi_c\,\psi_0^2\,\chi\right)}}{16\,\psi_0\,\phi_c}.
\end{align}
Using the $\delta$-N formalism, the scalar power spectrum is given by,
\begin{align}
\label{PsdeltaN}
    P_R=\dfrac{H^2}{4\,\pi^2}\left(N_{,\psi}^2+N_{,\phi}^2\right).
\end{align} 
Here, $N_{,\xi}=dN/d\xi$, $N_{,\psi}=N_{1,\psi}$ and $N_{,\phi}=N_{1,\phi}$ since the dominant contribution comes from phase-1 \cite{Clesse:2013jra}. Taking the derivative with respect to $\psi$ of \cref{efoldsph1} gives
\begin{align}
\label{Npsi}
N_{1,\psi}=\dfrac{-M^2\,\sqrt{m_1}\psi_0}{\psi_k\sqrt{b^{-2}\,M^4\,m_1+32\,\phi_c\,\psi_0^2\,\chi_2}}.
\end{align}
Here, $\chi_2\equiv\ln\left(\sqrt{\phi_c/m_1}M/2\,\psi_0\right)$ and $\psi_k\equiv d\psi/d\chi=\psi_0\,e^{\chi_k}$. One can calculate $\chi_k$ by solving \cref{xichi1} and \cref{xiefolds},
\begin{align}
    \chi_k=\dfrac{8\,\left(N_k-N_\text{e}\right)^2}{M^2 \,m_1\,\phi_c}-b_1^{-1}\,\left(N_k-N_\text{e}\right),
\end{align}
where, $N_k$, $N_\text{e}$ are the number of e-folds at the pivot scale, at the end of inflation respectively and we absorb the constant by redefining $b_1^{-1}\longrightarrow b^{-1}/\psi_0$. This reduces \cref{Npsi} to,
\begin{align}
\label{N1psi}
N_{1,\psi}=\dfrac{-M^2\,\sqrt{m_1}}{\sqrt{b^{-2}\,M^4\,m_1+32\,\phi_c\,\psi_0^2\,\chi_2}}\,\dfrac{1}{\exp{\left(\dfrac{8\,\phi_c\,\left(N_k-N_\text{e}\right)^2}{M^4\,m_1}-b_1^{-1}\,\left(N_k-N_\text{e}\right)\right)}}.
\end{align}
Similarly, from \cref{xiefolds} one can write $N_{1,\phi}$ as,
\begin{align}
\label{N1phi}
    N_{1,\phi} \simeq m_1\,\phi_c\,\dfrac{d\xi}{d\phi}\simeq -m_1.
\end{align}
The power spectrum \cref{PsdeltaN} now becomes,
\begin{align}
\label{PSanalyt}
    P_R \simeq \dfrac{H^2}{4\,\pi^2}\left(\dfrac{m_1^2}{m_\text{Pl}^4}+\left(\dfrac{\mathbb{b}\,M^2\,\sqrt{m_1}\,e^{-\left(\dfrac{8\,m_\text{Pl}^4\,\phi_c\left(N_k-N_\text{e}\right)^2}{M^4 \,m_1}-b^{-1}\,m_\text{Pl}\,\left(N_k-N_\text{e}\right)\right)}}{m_\text{Pl}^3\sqrt{b^{-2}\,M^4\,m_1+32\,\phi_c\,\psi_0^2\,\chi_2}}\right)^2\right).
\end{align}
Here, we introduce a fudge factor $\mathbb{b}=10^{-1}$ to keep the consistency of the previously used approximations and recover the Planckian units.
 In the left panel of \cref{fig:AnaExactPS_k_Fudge_factor}, we present an exact power spectrum (numerically solved) along with the analytical expression given in \cref{PSanalyt} for the potential \cref{potentialCl} and the BP in \cref{parmsetsCl}. This represents that the linear term helps to reduce the height of the power spectrum. The analytical expression is slightly off for the larger values of $b$ since estimations were made to obtain \cref{N1psi}. It is important to note that the analytical expression overestimates the power spectrum due to the fact that if we do not implement the assumption in \cref{xichi1}, the parameter $b^{-1}$ will evolve as a decreasing function due to the increasing behavior of the waterfall field. Therefore, the exponential factor would be suppressed. The width of the power spectrum is defined by the number of e-folds in the waterfall regime and hence is not affected by the underlying assumption.

As we know PBHs are formed in large over-dense regions, which means their formation rate can be significantly modified due to the non-Gaussian tails of the density perturbation distribution function as shown in several studies~\cite{Franciolini:2018vbk, Atal:2018neu, Ferrante:2022mui}.
However, investigating non-Gaussian effects remains an open problem, receiving a huge interest in current research. In our present analysis, we do not attempt to include the effects of non-Gaussianity; rather we consider them as uncertainties. Below we present a discussion regarding the impact of non-Gaussianities.
The first correction that arises if one goes beyond the Gaussian approximation is known as the scalar bi-spectrum $B_R$ which can be written as~\cite{Byrnes:2010ft, Planck:2015zfm} 
\begin{equation}\label{Bi}
B_R \equiv \left\langle R_{\bm{k}_{1}}R_{\bm{k}_{2}}R_{\bm{k}_{3}}\right\rangle=(2 \pi)^{3} \delta^{3}\left(\bm{k}_{1}+\bm{k}_{2}+\bm{k}_{3}\right) B_{R}\left(k_{1}, k_{2}, k_{3}\right),
\end{equation}
For detailed expressions of such bi-spectrum $B_R(k_1,k_2,k_3)$ see~\cite{Hazra:2012yn,Arroja:2011yj,Zhang:2020uek}.
Re-casting this in terms of the usual dimension-less $f_{\rm NL}$ parameter is expressed as \cite{Creminelli:2006rz,Byrnes:2010ft}
\begin{equation}\label{Fnl}
f_{\text{NL}}(k_1,k_2,k_3)=\frac{5}{6}\frac{B_{R}(k_1,k_2,k_3)}{P_{R}(k_1)
P_{R}(k_2)+P_{R}(k_2)P_{R}(k_3)+P_{R}(k_3)P_{R}(k_1)}.
\end{equation}
For the single field model, this implies the relation between spectral index and $n_s$ as $f_{\rm NL} |_{\rm squeezed} = 5 (1-n_s)/12$  in the squeezed limit $k_3\to 0$ and thereby $k_1=k_2$~\cite{Maldacena:2002vr, Creminelli:2004yq}.
The equilateral limit $k_1=k_2=k_3$ can be relevant for PBH formation for $P_R(k)$ which exhibits a peak, such that the PBH abundance gets approximately multiplied by~\cite{Franciolini:2018vbk, Zhang:2021vak}
\begin{equation}
    \exp\left[-\frac{23\,\delta_{\rm c}^3}{P_\zeta(k_{\rm peak})} f_{\rm NL}(k_{\rm peak},k_{\rm peak},k_{\rm peak})\right].
\end{equation}  
These considerations qualitatively suggest that unless $|f_{\rm NL}| \sim 1$, leading-order non-Gaussianities do not play a significant role in modifying PBH abundance. However, higher-order non-Gaussianities can be relevant with $R$ being expressed as a Gaussian fluctuation $R_{\rm G}$ in the local limit, such as those in ultra-slow roll inflationary models, see refs.~\cite{Atal:2019cdz, Ferrante:2022mui}.

Here we present an estimate of $f_{\rm NL}$ using
the $\delta N$ formalism following ref.~\cite{Clesse:2013jra}. The amplitude of scalar bi-spectrum is given by
\begin{align}
    f_\text{NL}=-\dfrac{5}{6}\dfrac{\sum_{i,j}N_{,i}N_{,j}N_{,ij}}{\left(\sum_iN_{,i}^2\right)^2}.
\end{align}
Here, $(i,j)$ refers to the derivative with respect to $(\phi, \psi)$. Considering phase 1, expanding the summation and using the chain rule, we arrive
\begin{align}
    f_\text{NL}=-\dfrac{5}{6}\dfrac{N_{1,\phi}^3+N_{1,\psi}^3}{\left(N_{1,\phi}^2+N_{1,\psi}^2\right)^2}\,\dfrac{\partial N_{1,\psi}}{\partial N}.
\end{align}
Using \cref{N1phi,N1psi}, it leads to the following analytical expression in $m_{\rm Pl}=1$ units,
\begin{align}
\label{fNL}
    f_\text{NL}&\simeq -\dfrac{5}{6}\dfrac{\exp{\left(b^{-1}\,(N_k-N_e)-\dfrac{8\,\phi_c\,(N_k-N_e)^2}{M^4\,m_1}\right)}\left(M^4\,m_1\,b^{-1}-16\,\phi_c\,(N_k-N_e)\right)}{m_1\left(M^5\exp{\left(2\,b^{-1}\,(N_k-N_e)-\dfrac{16\,\phi_c\,(N_k-N_e)^2}{M^4\,m_1}\right)}+M\,m_1\left(\dfrac{M^4m_1}{b^2}+32\,\phi_2\chi_2\psi_0^2\right)\right)^2}\\\notag
    &\times\left(M^6\exp{\left(3\,b^{-1}\,(N_k-N_e)-\dfrac{24\,\phi_c\,(N_k-N_e)^2}{M^4\,m_1}\right)}-m_1^{3/2}\left(\dfrac{M^4\,m_1}{b^2}+32\,\phi_2\chi_2\psi_0^2\right)^{3/2}\right).
\end{align}
The resulting $f_\text{NL}$ along with the power spectrum is presented in the right panel of \cref{fig:AnaExactPS_k_Fudge_factor} for the BP given in \cref{parmsetsCl}. We have shown the variation of $b$ as indicated in the inset legends. At the CMB scale $f_\text{NL}$ is very well consistent with the experimental bound i.e. $f_\text{NL}(\text{CMB})\lesssim 4.2$ \cite{Planck:2019kim}. The maximum predicted value is $f_\text{NL}\lesssim\pm 1$ which corresponds to the small scales.
Non-Gaussianities also somewhat modify the scalar-induced GWs as discussed in \cite{Ellis:2023oxs}.
Again, unless $f_{\rm NL} \sim 1$ which may give an order unity correction to their frequency spectrum, see e.g.~\cite{Ellis:2023oxs}, such effects are expected to be small. However, fully non-Gaussian contributions arising from higher orders are still a topic of active research and debate, with new contributions being identified which have been previously overlooked in the literature~\cite{Ellis:2023oxs, Perna:2024ehx, Li:2023qua}. A dedicated computation of the impact of such non-Gaussianities is beyond the scope of the current analysis and will be taken up in future studies.
\medskip

\section{Embedding in a Realistic Framework: Sneutrino Hybrid Inflation}
\label{realsiticframe}

In this section, we explore the SUSY imprints in the SIGWs and the PBHs.
We explore the supersymmetric hybrid inflation as a realistic model where the inflation is driven by a singlet sneutrino $\tilde{N}_R$. Consider the superpotential \cite{Antusch:2004hd},
\begin{align}
    \label{eq:W4CL}
 W = \kappa S \left(\frac{\Hat{\phi}^4}{{M'}^2} -
M_1^2\right) +
\frac{(\lambda_{N})}{M_\star}
\,\Hat{N}\Hat{N}\,\Hat{\phi}\Hat{\phi} \, 
+ \dots \,,
\end{align}
where $\kappa$ and $(\lambda_{N})$ are dimensionless Yukawa
couplings and ${M'},M_1$ and $M_\star$ are three independent mass parameters.  
The superfields $\Hat{N}$, $\Hat{\phi}$ and $\Hat{S}$ contain the bosonic components which are respectively:
the singlet sneutrino $\Hat{N}$ which plays the role of the inflaton;  
the waterfall field $\Hat{\phi}$, which is not exactly at zero during inflation but slightly off due to the presence of the linear term and develops a non-zero vacuum expectation value (vev) after inflation; 
and the singlet field $S$ is held at zero during and after inflation.
Note that we assume a $Z_4$ symmetry to prevent an explicit right-handed (s)neutrino mass for the superfield containing the inflaton. Any other RH neutrinos are assumed to be singlets under the $Z_4$ and have large explicit masses.

The vev of the waterfall field after inflation is fixed by the first term on the right-hand side in \cref{eq:W4CL}. During inflation, this term contributes a large vacuum energy to the potential. The waterfall superfield appears as $\Hat{\phi}^4/M^{'2}$ which will prevent the explicit singlet sneutrino masses via $Z_4$ discrete symmetry but this will be softly broken due to the presence of the linear term. The superpotential \cref{eq:W4CL} also respect the $U(1)_R$-symmetry and carries unit $R$-charge by assigning unit to $\Hat{S}$ and $1/2$ to $\Hat{N}$. 
The discrete subgroup of $Z_4$ symmetry acts as a matter parity under suitable conditions \cite{King:1997ia, Dvali:1997uq}. The K\"ahler potential with non-zero F-terms during inflation is given as 
\cite{Antusch:2004hd}
\begin{align}
    \label{noncanKahl}\mathrm{K}&=|\Hat{S}|^2+|\Hat{\phi}|^2+|\Hat{N}|^2+\kappa_S
\dfrac{|\Hat{S}|^4}{4\,m_\text{Pl}^2}+\kappa_N
\dfrac{|\Hat{N}|^4}{4\,m_\text{Pl}^2}+\kappa_\phi
\dfrac{|\Hat{\phi}|^4}{4\,m_\text{Pl}^2}+\kappa_{S\phi}
\dfrac{|\Hat{S}|^2|\Hat{\phi}|^2}{m_\text{Pl}^2}\\\notag
&+\kappa_{SN}
\dfrac{|\Hat{S}|^2|\Hat{N}|^2}{m_\text{Pl}^2}+\kappa_{N\phi}
\dfrac{|\Hat{N}|^2|\Hat{\phi}|^2}{m_\text{Pl}^2}+\dots
\end{align}
where the dots are for higher-order terms \footnote{\Cref{potentialCl} is just used as a toy model without committing to a particle theory scenario like SUSY, etc. The results obtained in this toy model help us study the (s)neutrino model presented in \cref{snpotenCL} based on SUSY. For instance, the terms containing dimension-6 in \cref{potentialCl} in the realistic SUSY model are allowed under the assumed $Z_4$-symmetry. Such terms in SUSY (and SUGRA) with the cut-off scale $M_{\star}$ are perfectly consistent as the model is assumed to be valid up to this energy scale and not beyond that. For more details of such discussion see the original model presented in \cite{Antusch:2004hd}.}  and $m_\text{Pl}=2.43\times 10^{18}$~GeV is the reduced Planck mass. The field $S$ acquires a large mass and sits at zero during inflation. 
The F-term scalar potential is given by \cite{Antusch:2004hd},
\begin{align}
    V_F=e^{\mathrm{K}/m_\text{Pl}^2}\left(K^{-1}_{ij} D_{z_i}\mathrm{W}D_{z_j^*}\mathrm{W}^*-3\dfrac{|\mathrm{W}|^2}{m_\text{Pl}^2}\right).
\end{align}
Here, $z_i\in\{\Hat{N},\Hat{\phi},\Hat{S}\dots\}$ are the bosonic components of the superfields and,
\begin{align}
    D_{zi}=\dfrac{\partial \mathrm{W}}{\partial z_i}+\dfrac{\mathrm{W}}{m_\text{Pl}^2}\dfrac{\partial\mathrm{K}}{\partial z_i},\,\,\,\,\,\,\,K_{ij}=\dfrac{\partial^2\mathrm{K}}{\partial z_i \partial z_j^*}, \,\,\,\,\,\text{and}\,\,\,\,\, D_{z_j^*}\mathrm{W}^{*}=(D_{z_j}\mathrm{W})^{*}.
\end{align}

Assuming that $\Tilde{N}, \Tilde{\phi}$ and $S$ are effective gauge singlets and there are no relevant D-terms at the considered energy scale. In terms of real fields, the scalar potential for sneutrino hybrid inflation close to the critical point of instability, using \cref{eq:W4CL,noncanKahl}. We also incorporate a soft-breaking of $Z_4$-symmetry with a linear term, which as we will show will inflate topological defects and help us to control PBH overproduction  \cite{Stamou:2024lqf,Afzal:2024xci} can be written as \cite{Antusch:2004hd},
\begin{align}
\label{snpotenCL}
V(\tilde{N}_R,\tilde{\phi})&\supseteq
\kappa^2\,M_1^4\left(1- \dfrac{\beta}{2}\left(\dfrac{\tilde{\phi}}{m_\text{Pl}}\right)^2-\left(\dfrac{1}{M_1\,M^{'}}\right)^2\,\tilde{\phi}^4+\dfrac{\gamma}{2\,m_\text{Pl}^2}\,\left(\tilde{N}_R-\tilde{N}_{Rc}\right)^2 \right. \\\notag
&\left.+c_1^3\,\left(\dfrac{\tilde{N}_R-\tilde{N}_{Rc}}{\kappa^2\,M_1^4}\right)+\dfrac{\lambda_N^2}{2\,M_\star^2\,\kappa^2\,M_1^4}\,\tilde{N}_R^4\,\tilde{\phi}^2+b_1^3\,\dfrac{\tilde{\phi}}{\kappa^2\,M_1^4}\right).
\end{align}
Here, $\beta$ and $\gamma$ are dimensionless couplings defined as \cite{Antusch:2004hd},
\begin{align}
    \beta = \kappa_{S\phi}-1,\,\,\,\,\,\,\,\,\,\,\,\, \gamma=1-\kappa_{SN}.
\end{align}
The coefficients of the linear terms $c_1$ and $b_1$ are dimensionful parameters and $b_1$ controls the peak of the scalar power spectrum to avoid PBH overproduction, as discussed in \cref{PBHOverpro,PBHOverpropsana}. The schematic view of the hybrid potential is shown in \cref{fig:potential}. 
\subsection{Comparing Model Parameters and Power Spectrum}
\label{sec:Modelcomp}
Let us now fix the BPs for \cref{snpotenCL} by comparing with the toy model \cref{potentialCl}, we identify,
\begin{align}
    \kappa^2\, M_1^4&=\Lambda,\,\,\,\,\,\,\,\,\,\,\ M_1^2\,M^{'2}=M^4,\,\,\,\,\,\,\,\,\,\,\ \dfrac{\beta}{2\,m_\text{Pl}^2}=\dfrac{2}{M^2},\,\,\,\,\,\,\,\,\,\,\ \dfrac{-\gamma}{2\,m_\text{Pl}^2}=\dfrac{1}{m_2^2} \\\notag
    &\dfrac{\lambda_N^2}{2\,M_\star^2\,\kappa^2\,M_1^4}=\dfrac{1}{M^2\,\phi_c^4},\,\,\,\,\,\,\,\,\,\,\ \dfrac{b_1^3}{\kappa^2\,M_1^4}=\dfrac{1}{b},\,\,\,\,\,\,\,\,\dfrac{c_1^3}{\kappa^2\,M_1^4}= \dfrac{1}{m_1}.
\end{align}
Following \cref{auxfield}, the auxiliary field distribution width can be written as,
\begin{align}
    \phi_0=\left(\dfrac{\kappa^2\,M_1^4}{92\,\pi^{3/2}\,m_\text{Pl}^4}\,\sqrt{\dfrac{2\,M_1\,M^{'}\,\tilde{N}_{R_c}}{c_1^3}}\right)^{1/2}.
\end{align}
The benchmark point for the potential \cref{snpotenCL} is given in \cref{parmsetsSNCL}. The abundance of PBH for the benchmark point in \cref{parmsetsSNCL} is shown in \cref{fig:f_pbhSNCL} that explains the PBHs as DM entirely or some fraction of it. The model predicts the scalar spectral index, $n_s$ and the tensor to scalar ratio, $r$ consistent with recent Planck 2018 results \cite{Planck:2018jri}. Following \cref{CLns,epetCL} we obtain,
\begin{align}
    n_s\simeq1+2\,\gamma,\,\,\,\,\,\,\,\, r= 16\,\epsilon_V\simeq\dfrac{m_\text{Pl}^2\,c_1^6}{2\,\Lambda^2}.
\end{align}
This fixes $\gamma=-0.017$ for the central value of $n_s$. The mass squared of the waterfall field at $\tilde{\phi}=0$ is
\begin{align}
M_{\tilde{\phi}}^2=\left(-\dfrac{\kappa^2\,M_1^4\,\beta}{m_\text{Pl}^2}+\dfrac{\lambda_N^2}{M_\star^2}\tilde{N}_R^4\right).
\end{align}
In this paper, we assume that $(\lambda_N/M_\star)^2 > \kappa^2\,\beta/(m_\text{Pl}^2)$ such that $M_{\tilde{\phi}}^2>0$ at large ${\tilde{N}_R}>{\tilde{N}_{Rc}}$ to stabilize the inflationary trajectory at $\tilde{\phi} = 0$, where,
\begin{align}
    \tilde{N}_{Rc}=\left(\dfrac{\kappa^2\,M_1^4\,M_\star^2\,\beta}{m_\text{Pl}^2\,\lambda_N^2}\right)^{1/4}.
\end{align}
During inflation, as long as $\tilde{N}_R\lesssim\tilde{N}_{Rc}$, the effective mass square of $\tilde{\phi}$ becomes negative, giving rise to tachyonic instability that will grow the curvature perturbations. These growing perturbations will enhance the scalar power spectrum at small scales and upon horizon re-entry the collapse of large density fluctuations produces the PBHs. When the waterfall field acquires a non-zero vev during the waterfall transition, the positive mass square term yields the masses of the singlet sneutrinos.  The benchmark points are given in \cref{parmsetsSNCL} for the potential in \cref{snpotenCL} for the production of PBH and induced SGWB.

\begin{figure}[t]
    \centering
    \includegraphics[width=0.75\linewidth]{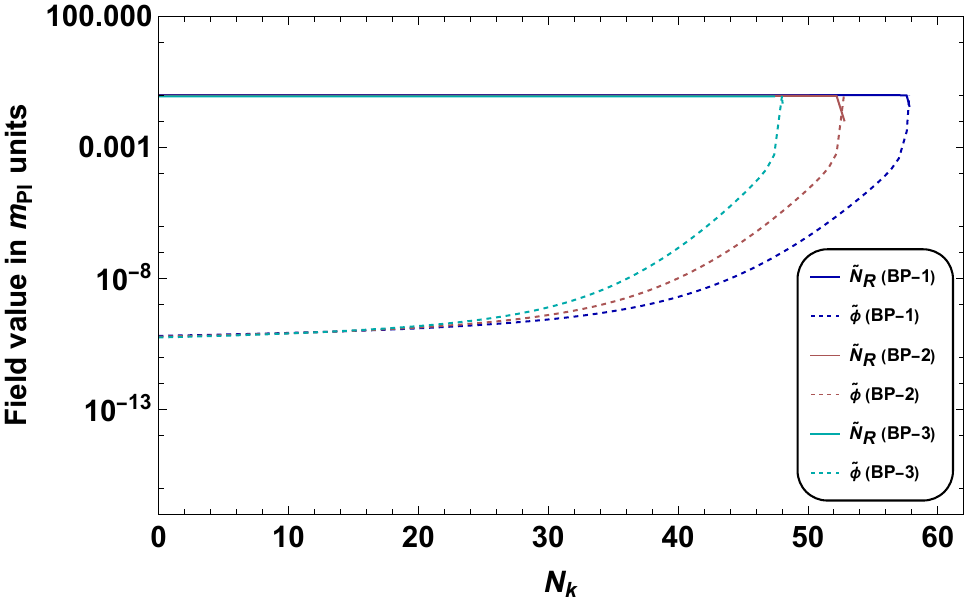}
    \caption{\it Fields evolution with the number of e-folds from pivot scale to the end of inflation. We evaluate solving the full background \cref{bac} using the potential given in \cref{snpotenCL} for the benchmark points in \cref{parmsetsSNCL}. }
    \label{fig:phipsiCl}
\end{figure}
\begin{figure}[t]
    \centering
    \includegraphics[width=0.8\linewidth]{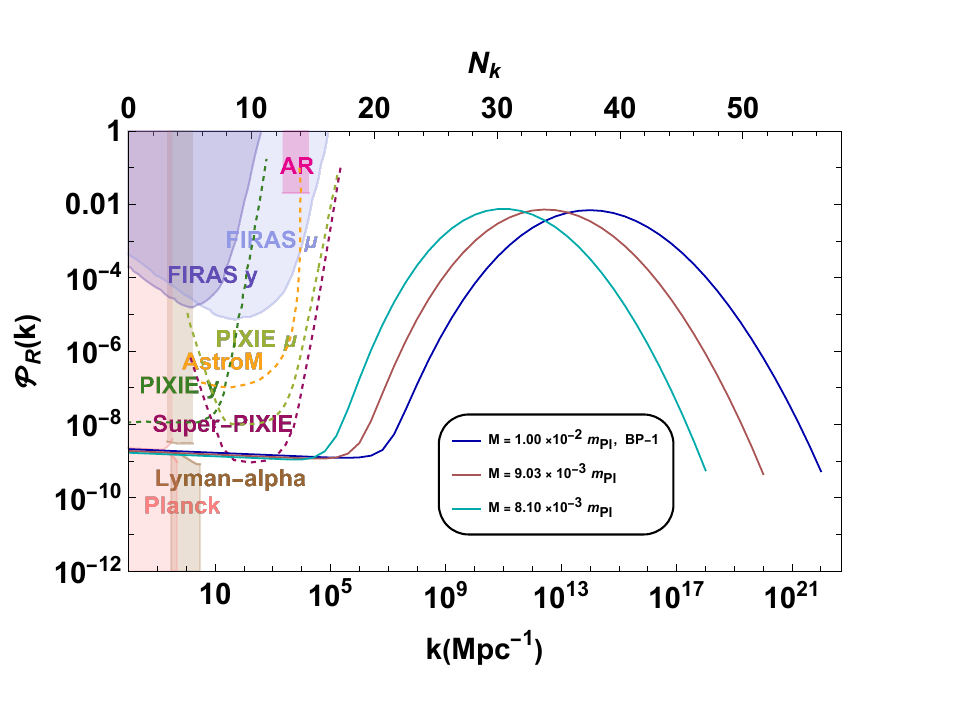}
    \caption{\it Power spectrum by solving the exact linear scalar perturbation equations \cref{Powspec} with the shaded region corresponding to the constraints from the present (solid) and future (dashed) experiments. The corresponding set of parameters are given in \cref{parmsetsSNCL}.}
    \label{fig:PS_k_Cl}
\end{figure}

The field evolution from the pivot scale to the end of inflation is given in \cref{fig:phipsiCl} for the BPs in \cref{parmsetsSNCL}. The exact power spectrum \cref{Powspec} is given in \cref{fig:PS_k_Cl} from the pivot scale till the end of inflation. The power spectrum is constrained by the angular resolution of current CMB measurements at scales $10^{-4}\lesssim k/\text{Mpc}^{-1}\lesssim 1$. However, inhomogeneities at these scales result in isotropic deviations from the usual blackbody spectrum and are known as spectral distortions \cite{Chluba:2012we}.
\begin{table}[tbh!]
    \fontsize{12pt}{20pt}
    \caption{\it Benchmark points for model$-1$ parameters} 
    \centering 
    \begin{adjustbox}{max width=\columnwidth}
        \begin{tabular}{|M{1.3cm} |M{1.9cm} |M{1.4cm}|M{2.2cm} |M{1.3cm} |M{2.2cm} |M{2.2cm}|M{2.5cm}|M{2.5cm}|}
            \hline
            \bf{Model}& \bf{$M_1/m_\text{Pl}$}&\bf{$M_\star/m_\text{Pl}$}&\bf{$-(M^{'}/m_\text{Pl})^2$}&\bf{$\beta$}&\bf{$\lambda_N$}&\bf{$\kappa$}&\bf{$b_1/m_\text{Pl}$}&\bf{$c_1/m_\text{Pl}$}\\ [0.6ex] 
            \hline\hline
            $\text{BP-1}$ &$1.00\times 10^{-2}$ &$1.0$ &$-1.0$ &$384.5$  & $1.73\times 10^{-6}$ & $ 8.50\times 10^{-6} $& $ -9.11\times 10^{-11} $ & $1.4\times 10^{-8}$\\
            \hline
            $\text{BP-2}$ &$9.02\times 10^{-3}$ &$1.0$ &$-1.0$ &$443.2$  & $2.14\times 10^{-6}$ & $ 1.13\times 10^{-5} $& $ -9.11\times 10^{-11} $ & $1.4\times 10^{-8}$\\
             \hline
            $\text{BP-3}$ &$8.10\times 10^{-3}$ &$1.0$ &$-1.0$ &$493.8$  & $2.52\times 10^{-6}$ & $ 1.40\times 10^{-5} $& $ -9.11\times 10^{-11} $ & $1.4\times 10^{-8}$\\
            \hline
        \end{tabular}
    \end{adjustbox}
    \label{parmsetsSNCL}
\end{table}
\begin{table}[tbh!]
    \fontsize{12pt}{20pt}
    \centering 
    \begin{adjustbox}{max width=\columnwidth}
        \begin{tabular}{|M{1.3cm} |M{1.9cm} |M{1.6cm}|M{2.5cm} |M{1.3cm} |M{1.3cm} |M{1.4cm}|M{2.5cm}|M{2.5cm}|M{1.3cm}|}
            \hline            \bf{Model}&\bf{$\gamma$}&\bf{$\tilde{N}_{Rc}/m_\text{Pl}$}&\bf{$\tilde{N}_{Ri}/m_\text{Pl}$}&\bf{$\tilde{\phi}_i/m_\text{Pl}$}&\bf{$N_k$}&\bf{$n_s$}&\bf{$r$}&\bf{$T_R/\text{GeV}$}&\bf{$\delta_c$}\\ [0.6ex] 
            \hline\hline
           $\text{BP-1}$ &$-0.017$ &  $0.100$ &$\tilde{N}_{Rc}(1+0.001)$ & $\phi_0$& $58$ & $0.966$ & $2.89\times 10^{-11}$&$5.6\times10^7$&$0.52$\\
           \hline
           $\text{BP-2}$ &$-0.017$ &  $0.095$ &$\tilde{N}_{Rc}(1+0.001)$ & $\phi_0$& $53$ & $0.966$ & $3.25\times 10^{-11}$&$6.3\times10^7$&$0.45$\\
            \hline
           $\text{BP-3}$ &$-0.017$ &  $0.090$ &$\tilde{N}_{Rc}(1+0.001)$ & $\phi_0$& $48$ & $0.966$ & $3.62\times 10^{-11}$&$6.8\times10^7$&$0.6$\\
            \hline
        \end{tabular}
    \end{adjustbox}
    \label{table1}
\end{table} 
There are two major categories of these distortions: $\mu$-distortions, associated with chemical potential that occurs at early times, and Compton $y$-distortions, generated at redshifts $z \lesssim 5 \times 10^4$. A $\mu$-distortion is associated with a Bose-Einstein distribution
with $\mu \neq 0$. Currently, the COBE/FIRAS instruments put the most stringent constraints on spectral distortions, which restricts $\lvert\mu\rvert \lesssim 9.0 \times 10^{-5}$ and $\lvert y \rvert \lesssim 1.5\times 10^{-5}$ at the $95\%$ C.L \cite{Fixsen:1996nj, Byrnes:2024vjt, Iovino:2024tyg}. A PIXIE-like detector can investigate distortions with magnitudes $\mu\lesssim 2\times 10^{-8}$ and $y\lesssim 4 \times 10^{-9}$ \cite{A_Kogut_2011}. The power spectrum on small scales, should be around $10^{-2}-10^{-3}$ to explain the maximum abundance of DM. While the peak is constrained from FIRAS on large scales see the bounds in \cref{fig:PS_k_Cl}. We find that our model parameters in \cref{parmsetsSNCL} satisfy the constraints of COBE/FIRAS. It is important to note that in the BPs \cref{parmsetsSNCL}, although the mass scale $M$ is sub-Planckian the coupling $\beta$ is such a large number. We will see later in \cref{sec:alphaattrac} that it can be controlled by introducing a field transformation by the so-called $\alpha$-attractor.

\smallskip
\section{PBH abundance}
\label{sec:PBHabund}
The mass of PBH formation is associated with a wave vector $k$ and is given by \cite{Ballesteros:2017fsr},
\begin{align}\label{Mrange}
	M_{\text{PBH}}=3.68\left(\dfrac{\gamma_c}{0.2}\right)\left(\dfrac{g_{*}(T_f)}{106.75}\right)^{-1/6}\left(\dfrac{10^6\,\text{Mpc}^{-1}}{k}\right)^2\,M_{\odot}.
\end{align}
The fractional abundance of PBHs, $\Omega_{\text{PBH}}/\Omega_{\text{DM}}\equiv f_{\text{PBH}}$ is defined as \cite{Ballesteros:2017fsr},
\begin{align}
	\label{abund}
 f_{\text{PBH}}=\dfrac{\beta(M_{\text{PBH}})}{3.94\times 10^{-9}}\left(\dfrac{g_{*}(T_f)}{106.75}\right)^{-1/4}\left(\dfrac{\gamma_c}{0.2}\right)^{1/2}\left(\dfrac{0.12}{\Omega_{\text{DM}} h^2}\right)\left(\dfrac{M_{\text{PBH}}}{M_{\odot}}\right)^{-1/2},
\end{align}
where $M_{\text{PBH}}$ is the PBH mass, the current energy density of DM is $h^2\Omega_{\text{DM}}=0.12$, $\gamma_c=0.2$ is the factor depends on the gravitational collapse and $\beta$ is the fractional energy density at the time of formation and is given by \cite{Motohashi:2017kbs},
\begin{align}
\label{beta}	\beta(M_{\text{PBH}})=\dfrac{1}{2\pi\sigma^2(M_{\text{PBH}})}\int_{\delta_c}^{\infty}d\delta\,\,\, \text{exp}\left(-\dfrac{\delta^2}{2\sigma^2\left(M_{\text{PBH}}\right)}\right).
\end{align}
 The variance, $\sigma(M_\text{PBH})$ of curvature perturbations ranges between $\sigma^2(M_\text{PBH})\simeq10^{-2}-10^{-3}$ which corresponds to the critical threshold, $\delta_c\simeq 0.4-0.6$ \cite{Musco:2020jjb,Escriva:2020tak,Escriva:2019phb,Musco:2018rwt} to explain the entire abundance of DM. Although $\delta_c$ is not a free parameter, it depends on both the shape of the power spectrum~\cite{Musco:2020jjb} and the choice of the window function~\cite{Young:2019osy}, still there remains some uncertainty in the literature regarding the calculation of the threshold, linked to super-horizon corrections~\cite{DeLuca:2023tun} and the average profile for the compaction function~\cite{Ianniccari:2024bkh}. In our analysis, we neglect the non-linear relationship between the density contrast and the curvature perturbation field and approximate the threshold without accounting for its dependence on the shape of the power spectrum and the window function; rather we have chosen it in such a way that it can explain the PBH as DM up to the allowed fraction in a given mass range. The corresponding $\delta_c$ values for each BPs are given in \cref{parmsetsSNCL} and \cref{parmsets}.
 Note that the parameter space is expected to change in a non-trivial way when moving beyond these approximations, particularly concerning corrections arising from non-linearities. \footnote{Note that there exists a non-linear relation between curvature perturbation and density contrast field that may have some impact on the PBH abundance which can be circumvented by the uncertainties in parameters like $\gamma_c$. We have discussed this in detail in the Appendix \cref{appen1}.} Mathematically, the variance is defined as \cite{Motohashi:2017kbs},
\begin{align}
	\sigma^2(M_\text{PBH}(k))=\dfrac{16}{81}\int \dfrac{dk^{'}}{k^{'}}(k^{'}/k)^4W^2(k^{'}/k) P_s(k^{'}),
\end{align} 
where $W(x)=\text{exp}(-x^2/2)$ is the Gaussian window function and $P_s(k)$ is the scalar power spectrum defined in \cref{Powspec}.

\begin{figure}[t]
    \centering
    \includegraphics[width=0.7\linewidth]{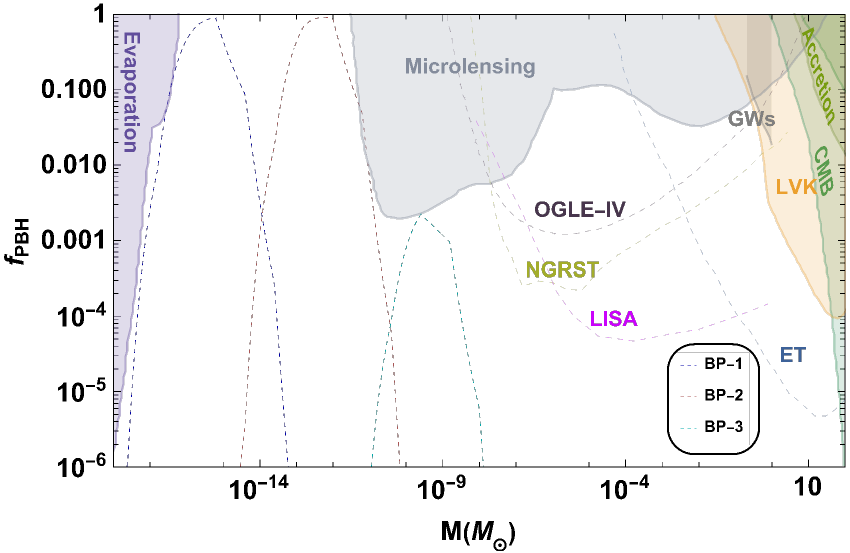}
\caption{\it PBH abundance as DM given in \cref{abund}. The shaded regions represent the
observational constraints on the PBH abundance from various experiments, see the main text
for the details.}
    \label{fig:f_pbhSNCL}
\end{figure}
In \cref{fig:f_pbhSNCL}, the PBH abundance \cref{abund} is demonstrated along with the different experimental constraints \cite{Green:2020jor} for given parameter sets in \cref{parmsetsSNCL}. Note that we consider the case of monochromatic mass functions. The bounds may change in \cref{fig:f_pbhSNCL} and \cref{fig:f_pbh} for an extended mass function \cite{Carr:2017jsz}. For BP-$1$, the entire abundance of DM can be explained by PBH.
The Hawking radiations may evaporate the PBHs and therefore there are some constraints in microlensing-related observations 
 such as: CMB\,\cite{Clark:2016nst}, EDGES\,\cite{Mittal:2021egv},  INTEGRAL\,\cite{Laha:2020ivk,Berteaud:2022tws}, Voyager\,\cite{Boudaud:2018hqb}, 511\;keV\,\cite{DeRocco:2019fjq},
EGRB\,\cite{Carr:2009jm}; HSC (Hyper-Supreme Cam)\,\cite{Niikura:2017zjd}, EROS\,\cite{EROS-2:2006ryy}, OGLE-IV \cite{Mroz:2024wia, Mroz:2024wag} and Icarus\,\cite{Oguri:2017ock};  if PBHs accrete, there are constraints due to the CMB spectrum ref.~\cite{Serpico:2020ehh}; finally the range around $M_{\odot}$ is constrained by LIGO-VIRGO-KAGRA observations on PBH-PBH merger \,\cite{Franciolini:2022tfm, DeLuca:2021wjr, Andres-Carcasona:2024wqk}). Future planned GW interferometers like LISA, Einstein Telescope (ET) and Cosmic Explorer (CE) are also expected to set limits on the PBH abundance see refs.~\cite{Kuhnel:2018mlr,DeLuca:2021hde,Franciolini:2023opt}, these are shown in dashed lines in the plot. Future sensitivity reaches of the Nancy Roman Telescope from micro-lensing are also presented, see ref.~\cite{DeRocco:2023gde}\footnote{A recent study of the sensitivity of X-ray telescopes studying x-ray pulsars to microlensing by sub-atomic size primordial black holes is done in~\cite{Tamta:2024pow}.}.

\section{Scalar-induced GWs}
\label{sec:SIGW}

\begin{figure}[t]
    \centering
    \includegraphics[width=0.8\linewidth]{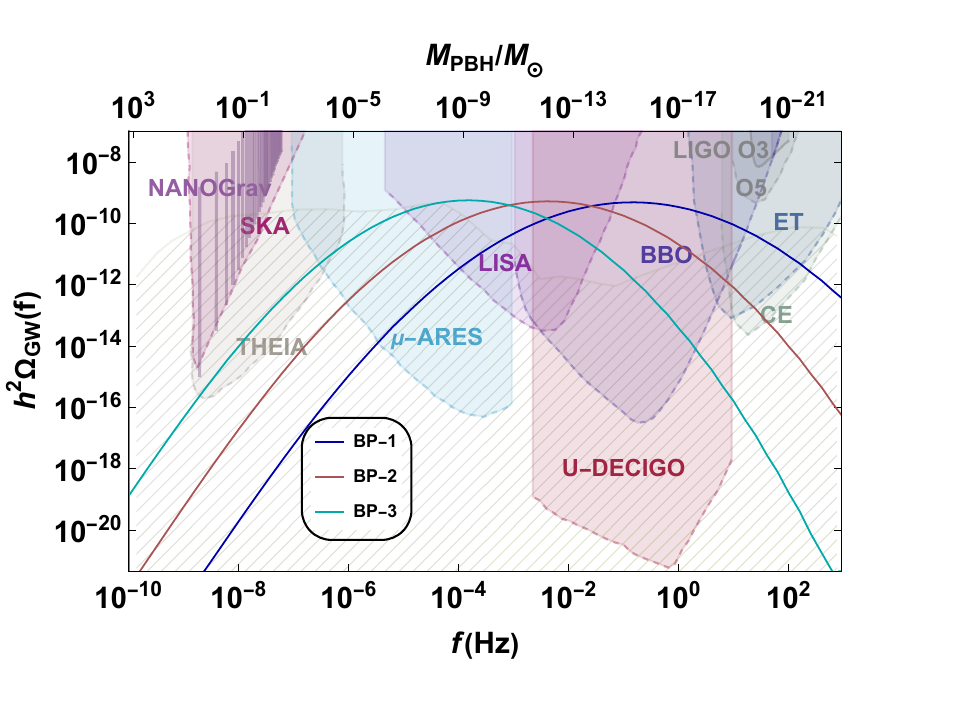}
\caption{\it The energy density of gravitational waves for \cref{Energdens} for the BPs in \cref{parmsetsSNCL}. The colored shaded regions indicate the sensitivity curves of present (solid boundaries) LIGO O3 \cite{KAGRA:2021kbb}, NANOGrav \cite{NANOGrav:2023gor} and future (dashed boundaries) LIGO O5, SKA \cite{Smits:2008cf}, THEIA \cite{Garcia-Bellido:2021zgu}, LISA \cite{Baker:2019nia}, $\mu$-ARES \cite{Sesana:2019vho}, BBO \cite{Corbin:2005ny}, U-DECIGO \cite{Yagi:2011wg, Kawamura:2020pcg}, CE \cite{Reitze:2019iox} and ET \cite{Punturo:2010zz} experiments. The hatched region shows the astrophysical background 
\cite{ghoshal2023traversing}.}
    \label{fig:OmegafSNCl}
\end{figure}

We assume that the formation of PBH in the radiation-dominated era, the energy density of GWs today in terms of scalar power spectrum \cref{Powspec}, is given by \cite{ Espinosa:2018eve, DeLuca:2020agl},
\begin{align}\label{Energdens}
	\Omega_{\text{GW}}(k)&=\dfrac{c_g\,\Omega_{r}}{6}\left(\dfrac{g_{*}(T_f)}{106.75}\right)\int_{-1}^{1}dd\int_{1}^{\infty}ds\,\,P_s\left(k\dfrac{s-d}{2}\right) P_s\left(k\dfrac{s+d}{2}\right) I(d,s), \\ \notag
	I(d,s)&=\dfrac{288(d^2-1)^2(s^2-1)^2(s^2+d^2-6)^2}{(d-s)^8(d+s)^8}\left\{\left(d^2-s^2+\dfrac{d^2+s^2-6}{2}\text{ln}\left|\dfrac{s^2-3}{d^2-3}\right|\right)^2\right.\\\notag
	&\left.\dfrac{\pi^2}{4}(d^2+s^2-6)^2\Theta(s-\sqrt{3}))\right\}.
\end{align}
Here, $\Omega_r=5.4\times10^{-5}$ is the present-day energy density of the radiation, $c_g=0.4$ in the SM, $\Theta$ is the Heaviside function, and $g_{*}(T_f)\simeq 106.75$ is the effective degrees of freedom at the temperature $T_f$ of PBH formation for SM like spectrum. Furthermore, using $k=2\pi f$, $1\text{Mpc}^{-1}=0.97154\times10^{-14}\,\text{s}^{-1}$ and $h=0.68$, we move into the $h^2\Omega_{\text{GW}} (f)-f$ plane. The GW spectra for the benchmark points in \cref{parmsetsSNCL} are shown in \cref{fig:OmegafSNCl} with the different experiments presented by the shaded region such as, SKA \cite{Smits:2008cf}, THEIA \cite{Garcia-Bellido:2021zgu}, LISA \cite{Baker:2019nia}, $\mu$-ARES \cite{Sesana:2019vho}, BBO \cite{Corbin:2005ny}, U-DECIGO \cite{Yagi:2011wg, Kawamura:2020pcg}, CE \cite{Reitze:2019iox} and ET \cite{Punturo:2010zz}.

 Searching for stochastic GW of cosmic origin is likely to reveal multiple astrophysical sources of GW background. They can primarily take the form of binary neutron star (NS-NS) events \cite{TheLIGOScientific:2017qsa} and LIGO/VIRGO detected binary black hole (BH-BH) merging events \cite{Abbott:2017oio, LIGOScientific:2018mvr}.
To differentiate with the scalar-induced GWs of cosmic origin, the foreground NS and BH can be subtracted using the sensitivities of the BBO and ET / CE windows, especially in the range $\Omega_{\rm GW} \simeq 10^{-15}$ \cite{Cutler:2005qq} and $\Omega_{\rm GW} \simeq 10^{-13}$ \cite{Regimbau:2016ike}.
The binary white dwarf galactic and extra-galactic (WD-WD) can be removed \cite{Kosenko:1998mv} with the expected sensitivity at $\Omega_{\rm GW} \simeq 10^{-13}$ \cite{Adams:2013qma}, and may be more significant than the NS-NS and BH-BH foregrounds in the LISA window \cite{Moore:2014lga}. The GW spectrum generated by the astrophysical foreground grew with frequency $\propto f^{2/3}$ in addition to this subtraction \cite{Zhu:2012xw}. This GW spectrum differs from that produced by second-order gravitational waves, which at low frequencies convict as $f^{3/2}$ and at higher frequencies as $f^{-3/2}$. This will allow us to identify the GW signals generated by scalar-induced sources.

\medskip

\section{{\it Model$-2$}: $\alpha$-attractor sneutrino hybrid inflation}
\label{sec:alphaattrac}

In this section, we explore the supersymmetric hybrid inflation in the context of an exponential $\alpha$-attractor model \cite{Kallosh:2022ggf}. As we have mentioned the predictions for the coupling $\beta$ is a very large number see \cref{parmsetsSNCL} in the previous sneutrino model. So, here we propose a variation of the sneutrino model where we will see that with the field redefinition in terms of $\alpha$-attractor, this parameter takes a natural value.
In terms of real fields, the scalar potential for sneutrino hybrid inflation (including a linear term) can be written as \cite{Antusch:2004hd},

\begin{align}\label{snpotenalpha}
V(\tilde{N}_R,\tilde{\phi})&\supseteq
\kappa^2\,M^4- \dfrac{\kappa^2\,M^4\,\beta}{2}\left(\dfrac{\tilde{\phi}}{m_\text{Pl}}\right)^2-\kappa^2\,\left(\dfrac{M}{M^{'}}\right)^2\,\tilde{\phi}^4+\dfrac{\kappa^2\,M^4\,\gamma}{2}\,\left(\dfrac{\tilde{N}_R}{m_\text{Pl}}\right)^2 \\\notag
&+\dfrac{\lambda_N^2}{2\,M_\star^2}\,\tilde{N}_R^4\,\tilde{\phi}^2+d^3\,\tilde{\phi}
\end{align}
with all the parameters as defined for \cref{snpotenCL}. The coefficient of the linear term $d$ is a dimensionful parameter that will again allow us to control the peak of the scalar power spectrum to avoid PBH overproduction. The canonical transformation of the inflaton field $\tilde{N}_R\longrightarrow \sqrt{6\,\alpha}\, \text{Tanh}\left(\tilde{\upsilon}_R/\sqrt{6\,\alpha}\right)$, allows us to write \cref{snpotenalpha} as,

\begin{align}
\label{snpotencanon}
V(\tilde{\upsilon}_R,\tilde{\phi})&= \kappa^2\,M^4- \dfrac{\kappa^2\,M^4\,\beta}{2}\left(\dfrac{\tilde{\phi}}{m_\text{Pl}}\right)^2-\kappa^2\,\left(\dfrac{M}{M^{'}}\right)^2\,\tilde{\phi}^4+\dfrac{\kappa^2\,M^4\,\gamma}{2}\,\left(\dfrac{\sqrt{6\,\alpha}\, \text{Tanh}\left(\tilde{\upsilon}_R/\sqrt{6\,\alpha}\right)}{m_\text{Pl}}\right)^2 \\\notag
&+\dfrac{\lambda_N^2}{2\,M_\star^2}\,\left(\sqrt{6\,\alpha}\, \text{Tanh}\left(\dfrac{\tilde{\upsilon}_R}{\sqrt{6\,\alpha}}\right)\right)^4\,\tilde{\phi}^2+d^3\,\tilde{\phi}.
\end{align}
The benchmark point for potential \cref{snpotencanon} is given in \cref{parmsets}. The PBH abundance as DM for the benchmark point in \cref{parmsets} is shown in \cref{fig:f_pbh}. The model predicts the scalar spectral index, $n_s$ and the tensor to scalar ratio, $r$ consistent with recent Planck 2018 results \cite{Planck:2018jri}.
The squared mass of the waterfall field at $\tilde{\phi}=0$ is,
\begin{align}
M_{\tilde{\phi}}^2=\left(-\dfrac{\kappa^2\,M^4\,\beta}{2\,m_\text{Pl}^2}+\dfrac{\lambda_N^2}{M_\star^2}\left(\sqrt{6\,\alpha}\,\text{Tanh}\left(\dfrac{\tilde{\upsilon}_R}{\sqrt{6\,\alpha}}\right)\right)^4\right).
\end{align}
Assuming $(\lambda_N\,6\,\alpha/M_\star)^2 > \kappa^2\,M^4\,\beta/(2\,m_\text{Pl}^2)$ such that $M_{\tilde{\phi}}^2>0$ at large ${\tilde{\upsilon}_R}>{\tilde{\upsilon}_{Rc}}$ to stabilize the inflationary trajectory at $\tilde{\phi} = 0$, where,
\begin{align}
    \text{Tanh}^4\left(\dfrac{\tilde{\upsilon}_{Rc}}{\sqrt{6\,\alpha}}\right)=\dfrac{\kappa^2\,M^4\,M_\star^2\,\beta}{2\,m_\text{Pl}^2\,\lambda_N^2\,(\sqrt{6\,\alpha})^4}.
\end{align}
\begin{figure}[H]
    \centering
    \includegraphics[width=0.7\linewidth]{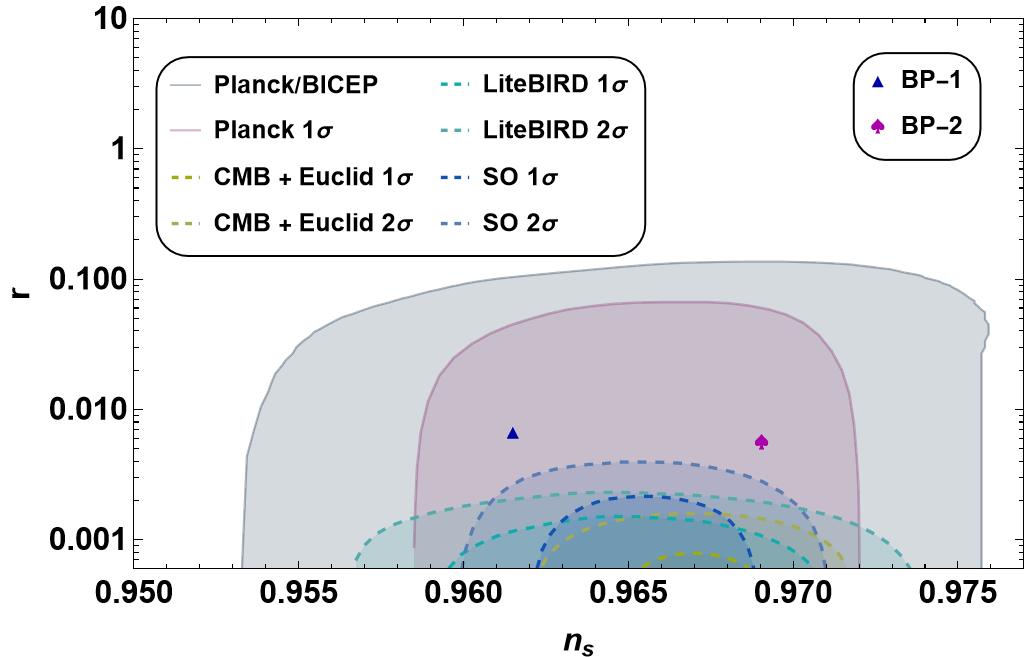}
    \caption{\it \it Tensor-to-scalar ratio $r$ vs. scalar spectral index $n_s$ for the corresponding parameter sets given in \cref{parmsets}. The solid contours are the current Planck bounds \cite{Planck:2018jri}, Planck/BICEP \cite{BICEP:2021xfz, Planck:2018vyg,BICEPKeck:2022mhb} and the dashed shaded region indicates the future proposed experiments (LiteBIRD, CMB-Euclid, Simons Observatory (SO)) \cite{laureijs2011euclid, SimonsObservatory:2018koc, LiteBIRD:2020khw}.}
    \label{fig:rnsalpha}
\end{figure}
\begin{figure}[t]
    \centering
    \includegraphics[width=0.7\linewidth]{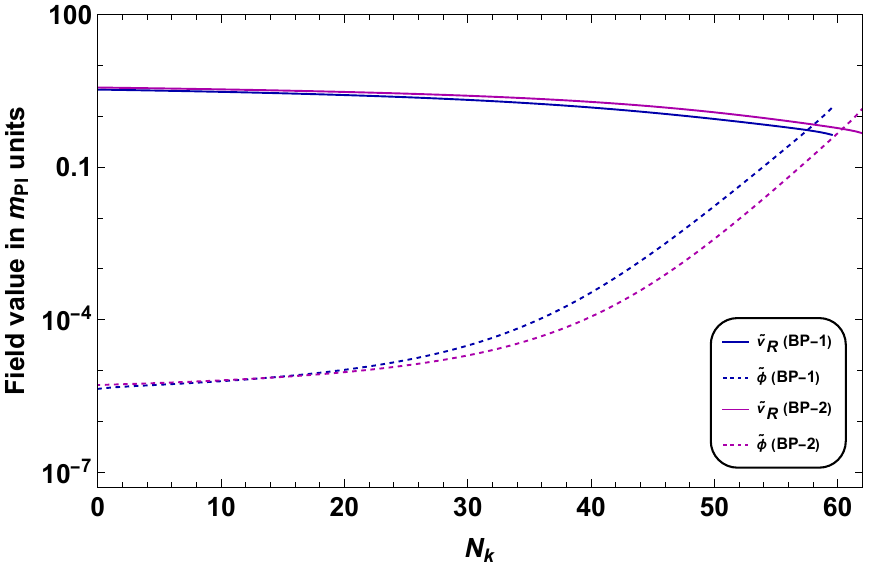}
    \caption{\it \it Fields evolution with the number of e-folds from pivot scale to the end of inflation. We evaluate solving the full background \cref{bac} using potential in \cref{snpotencanon} for the benchmark points in \cref{parmsets}. }
    \label{fig:phipsi}
\end{figure}

First we discuss what happens along the valley: as long as $\tilde{\upsilon}_R\lesssim\tilde{\upsilon}_{Rc}$, the effective mass square of $\tilde{\phi}$ stays negative. This gives rise to kind of tachyonic instability that leads to the large growth the curvature perturbations which consequently further enhance the scalar power spectrum at small scales of the universe. The collapse of large density fluctuations upon horizon reentry leads to PBH production. The presence of the linear term in the potential \cref{snpotencanon} serves as not to keep the field $\tilde{\phi}$ exactly at $\tilde{\phi}=0$ but makes it displaced. This also depends upon the sign of the coefficient of the linear term $d$. On the one hand, this inflates away unnecessary topological defects and on the other hand also controls the peak of the power spectrum at small scales to rescues from PBH overproduction. We take $M\sim O(1)$ to avoid eternal inflation \cite{Braglia:2022phb}. The parameter $\gamma$ controls the amplitude of the plateau in the valley, the coupling $\lambda_N$ defines the number of e-folds in the waterfall regime and $\kappa$ will fix the amplitude of the power spectrum at the pivot scale $k_\star=0.05\,\text{Mpc}^{-1}$ that is $A_s\simeq 2.24\times10^{-9}$. We consider the model involving exponential $\alpha$-attractor \cite{Kallosh:2022ggf} in order to to get rid of the problem of initial conditions that we usually have in a standard hybrid scenario \cite{Clesse:2015wea} (see \cite{Braglia:2022phb, Afzal:2024xci} for a detailed discussion of such initial conditions). Having considered these constraints, we define the BPs in \cref{parmsets} for the potential in \cref{snpotencanon}.
\begin{table}[t]
    \fontsize{12pt}{20pt}
    \caption{\it Benchmark points for model$-2$ parameters} 
    \centering 
    \begin{adjustbox}{max width=\columnwidth}
        \begin{tabular}{|M{1.3cm} |M{1.4cm} |M{1.6cm}|M{2.2cm} |M{1.3cm} |M{2.2cm} |M{2cm}|M{2cm}|}
            \hline
            \bf{Model}& \bf{$M/m_\text{Pl}$}&\bf{$M_\star/m_\text{Pl}$}&\bf{$-(M^{'}/m_\text{Pl})^2$}&\bf{$\beta$}&\bf{$\lambda_N$}&\bf{$\kappa$}&\bf{$d/m_\text{Pl}$}\\ [0.6ex] 
            \hline\hline
            $\text{BP-1}$ &$1.352$ &$1.0$ &$-29.246$ &$0.547$  & $7.07\times 10^{-7}$ & $ 1.7\times 10^{-6} $& $-3.0\times 10^{-6}$ \\
            \hline
               $\text{BP-2}$ &$1.352$ &$1.0$ &$-29.246$ &$0.600$  & $6.26\times 10^{-7}$ & $ 1.6\times 10^{-6} $& $-2.8\times 10^{-6}$ \\
            \hline
        \end{tabular}
    \end{adjustbox}
    \label{parmsets}
\end{table}
\begin{table}[t]
    \fontsize{12pt}{20pt}
    \centering 
    \begin{adjustbox}{max width=\columnwidth}
        \begin{tabular}{|M{1.3cm} |M{1.4cm} |M{1.6cm}|M{2.2cm} |M{1.3cm} |M{1.3cm} |M{1.5cm}|M{1.5cm}|M{2cm}|M{1.3cm}|}
            \hline
            \bf{Model}&\bf{$\gamma$}&\bf{$\sqrt{\alpha}/m_\text{Pl}$}&\bf{$\tilde{\upsilon}_{Ri}/m_\text{Pl}$}&\bf{$\tilde{\phi}_i/m_\text{Pl}$}&\bf{$N_k$}&\bf{$n_s$}&\bf{$r$}&\bf{$T_R/\text{GeV}$}&\textbf{$\delta_c$}\\ [0.6ex] 
            \hline\hline
           $\text{BP-1}$ &$0.073$ &$1$ & $3.5$ & $0$& $60$ & $0.961$ & $0.0068$&$9.8\times10^8$&$0.524$\\
            \hline
            $\text{BP-2}$ &$0.085$ &$1$ & $4.0$ & $0$& $62$ & $0.969$ & $0.0056$&$9.2\times10^8$&$0.460$\\
            \hline
        \end{tabular}
    \end{adjustbox}
    \label{table2}
\end{table}

\begin{figure}[tbh!]
    \centering
    \includegraphics[width=0.48\linewidth,height=5.4cm]{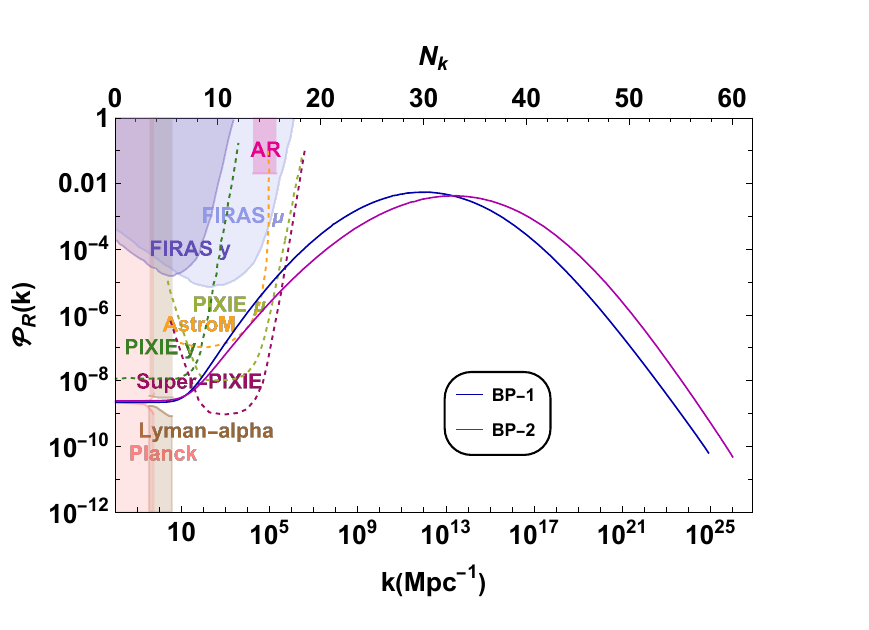}
    \quad
    \includegraphics[width=0.48\linewidth,height=5.4cm]{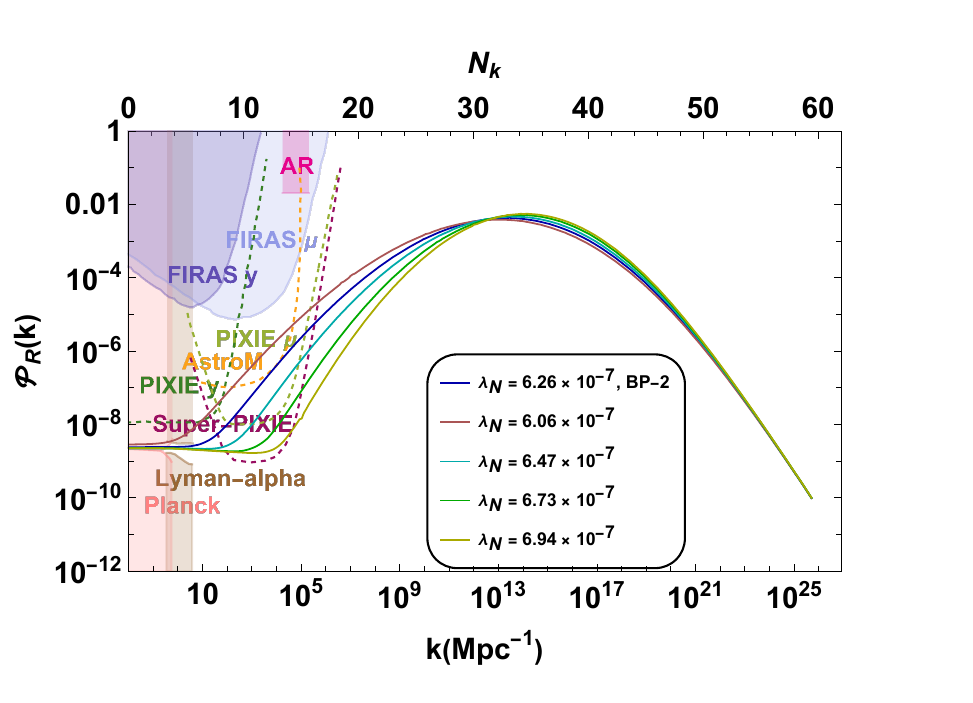}
    \quad
    \includegraphics[width=0.48\linewidth,height=5.4cm]{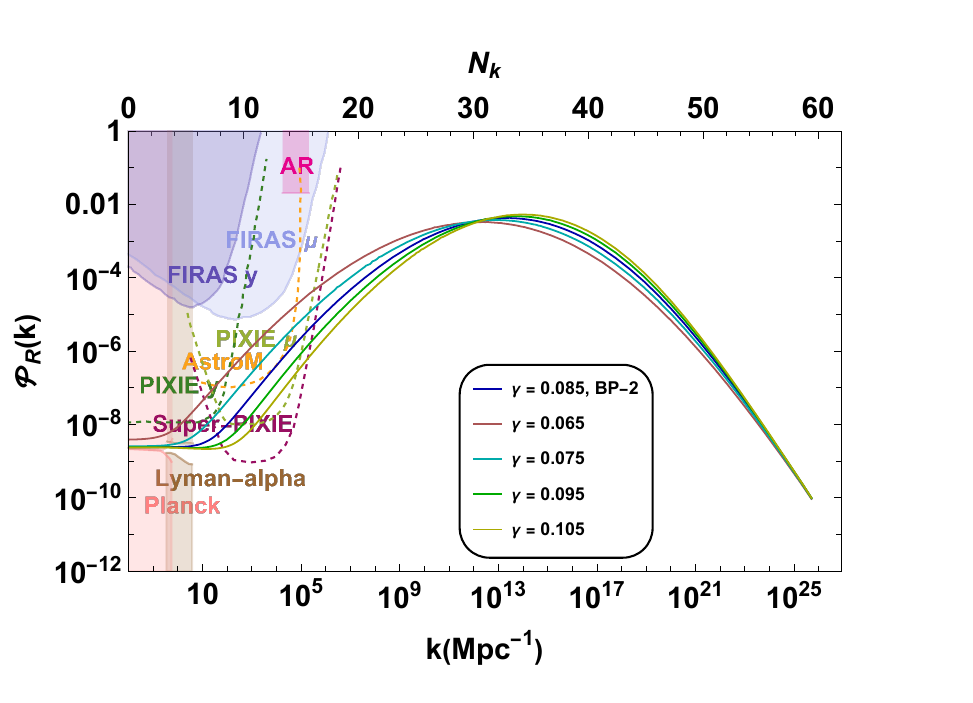}
    \quad
    \includegraphics[width=0.48\linewidth,height=5.4cm]{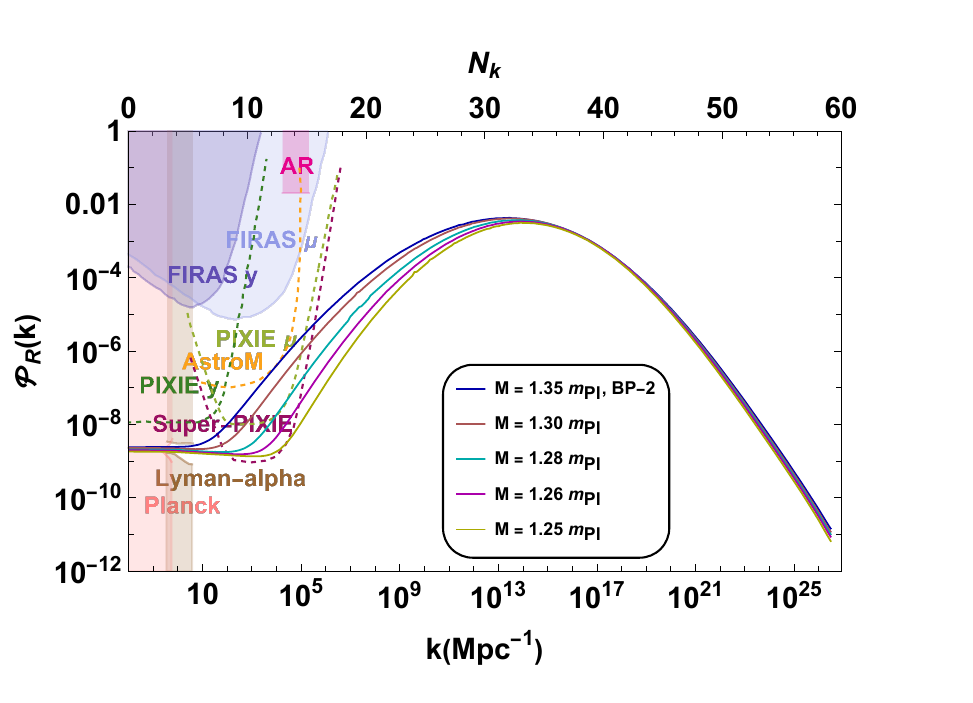}
    \quad
    \includegraphics[width=0.48\linewidth,height=5.4cm]{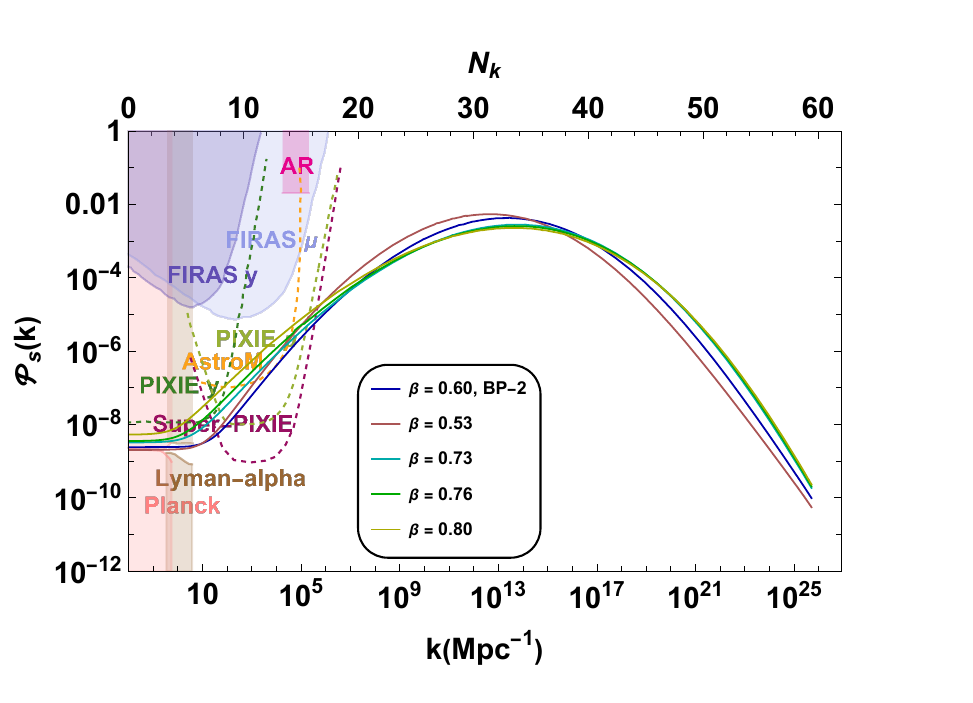}
     \quad
    \includegraphics[width=0.48\linewidth,height=5.4cm]{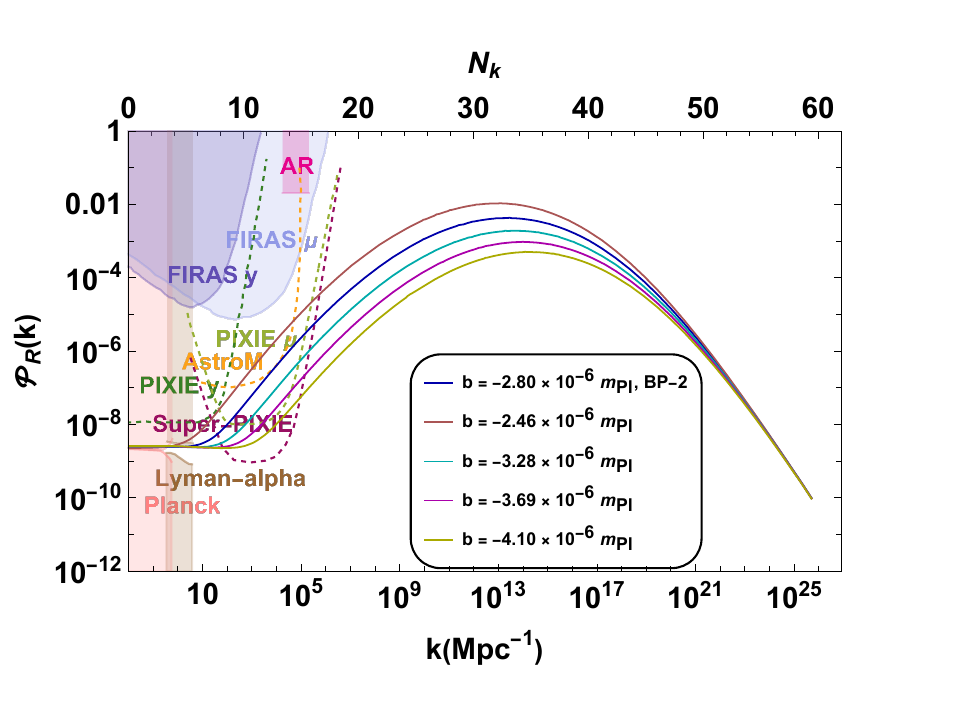}
    \caption{\it Power spectrum by solving the exact linear scalar perturbation equations with the shaded region corresponding to the constraints from present (solid) and future (dashed) experiments. The corresponding set of parameters is given in \cref{parmsets}.}
    \label{fig:PS_k}
\end{figure}

\begin{figure}[H]
    \centering
    \includegraphics[width=0.582\linewidth]{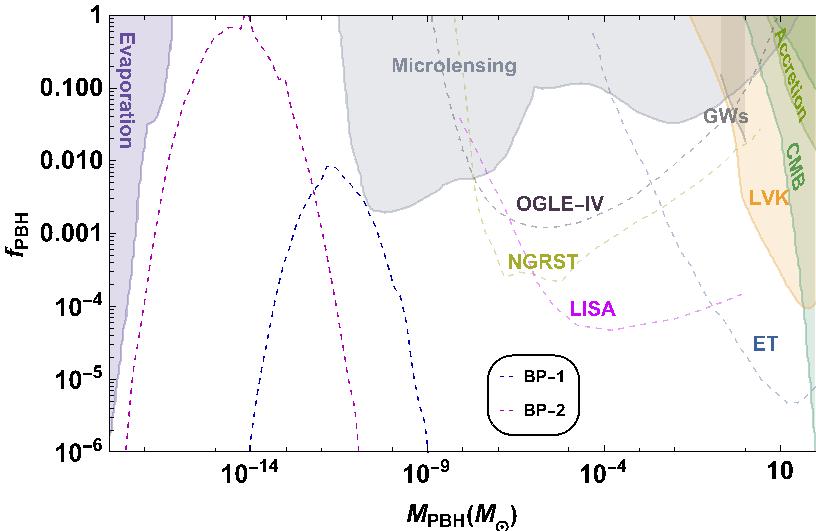}
\caption{\it PBH abundance given in \cref{abund} for the BPs \cref{parmsets}. The shaded regions represent the
observational constraints on the PBH abundance from various experiments, see the main text
for the details. For BP-1, as we see it can be the entire DM candidate of the universe.}
    \label{fig:f_pbh}
\end{figure}

\begin{figure}[t]
    \centering
    \includegraphics[width=0.484\linewidth,height=5.7cm]{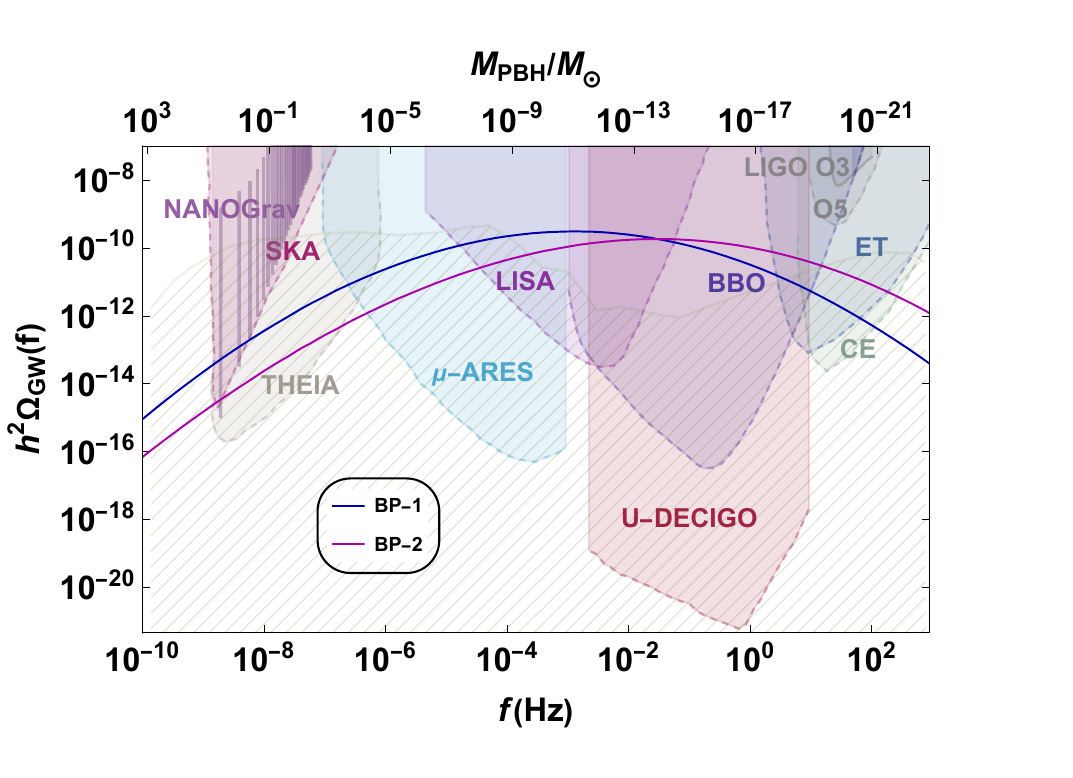}
     \quad
    \includegraphics[width=0.484\linewidth,height=5.7cm]{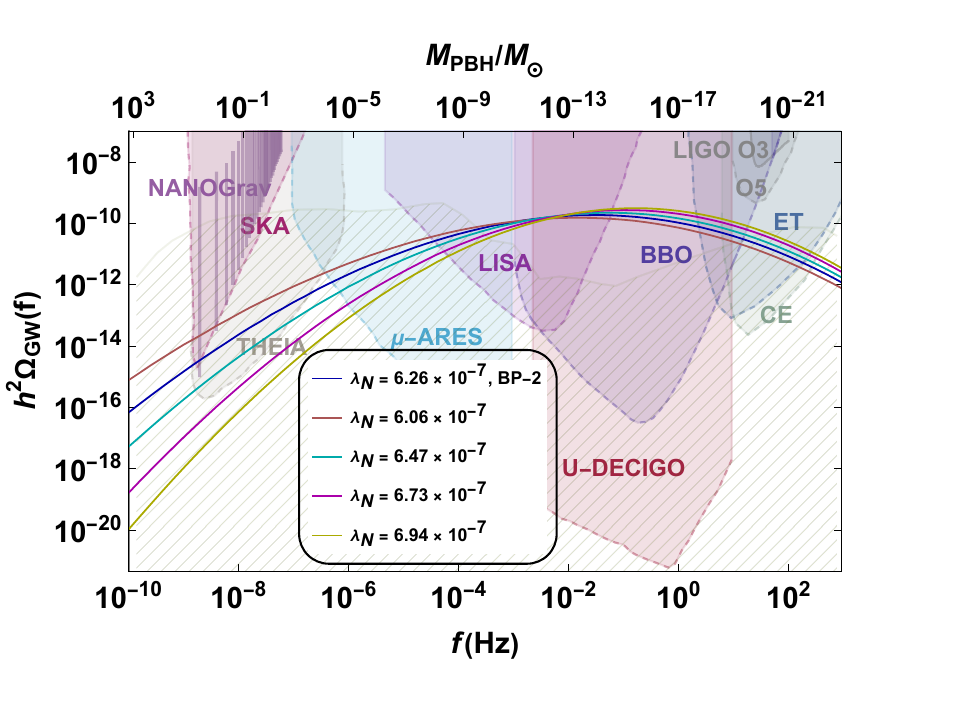}
    \quad
    \includegraphics[width=0.484\linewidth,height=5.7cm]{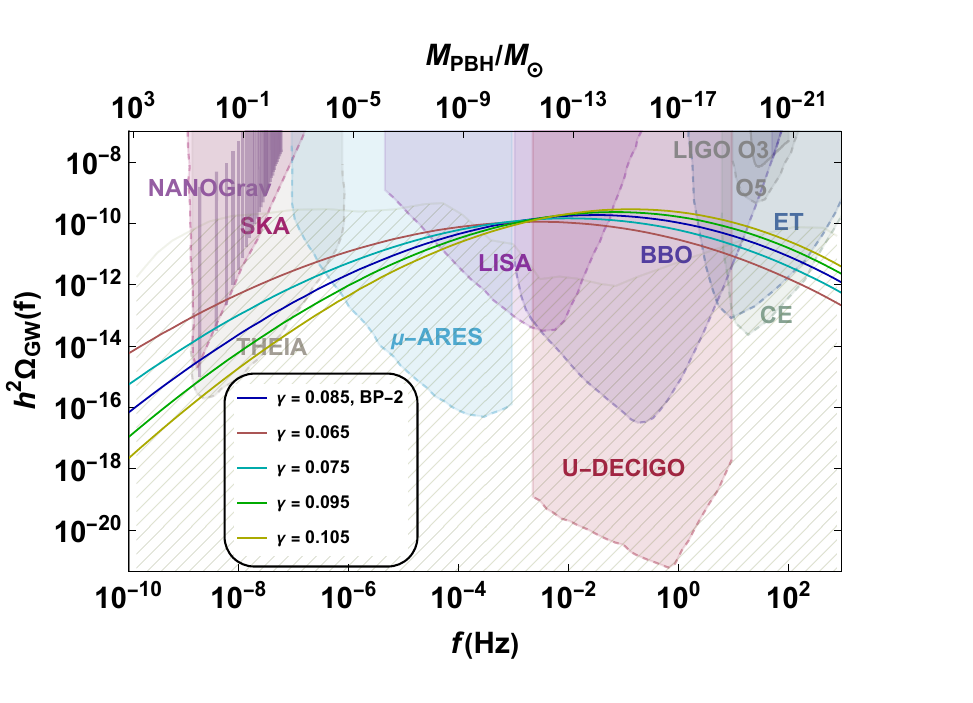}
    \quad
    \includegraphics[width=0.484\linewidth,height=5.7cm]{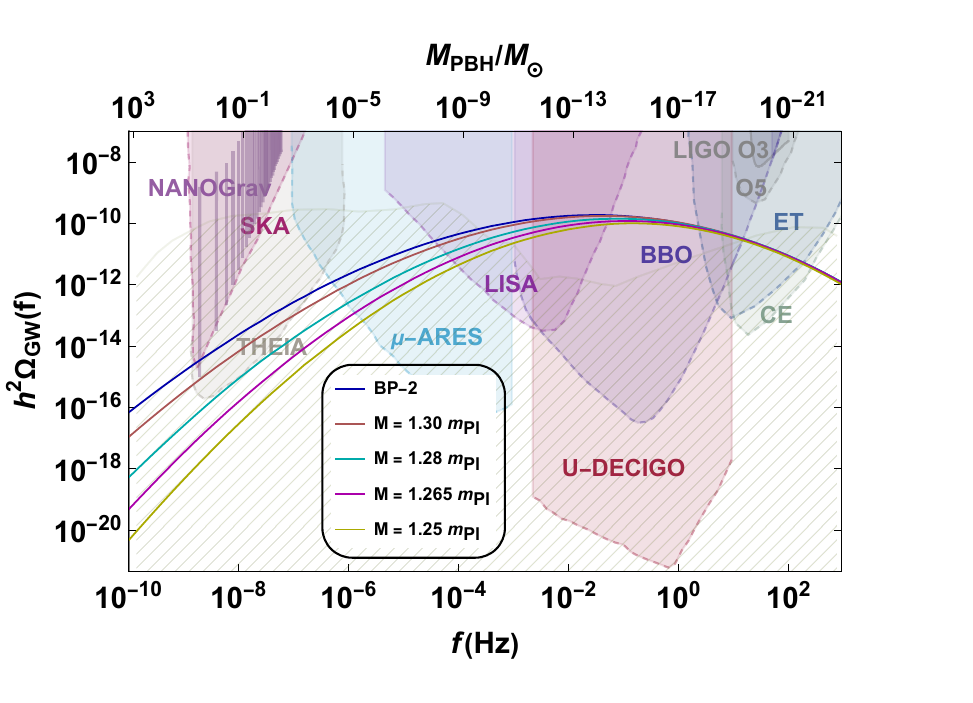}
    \quad
    \includegraphics[width=0.484\linewidth,height=5.7cm]{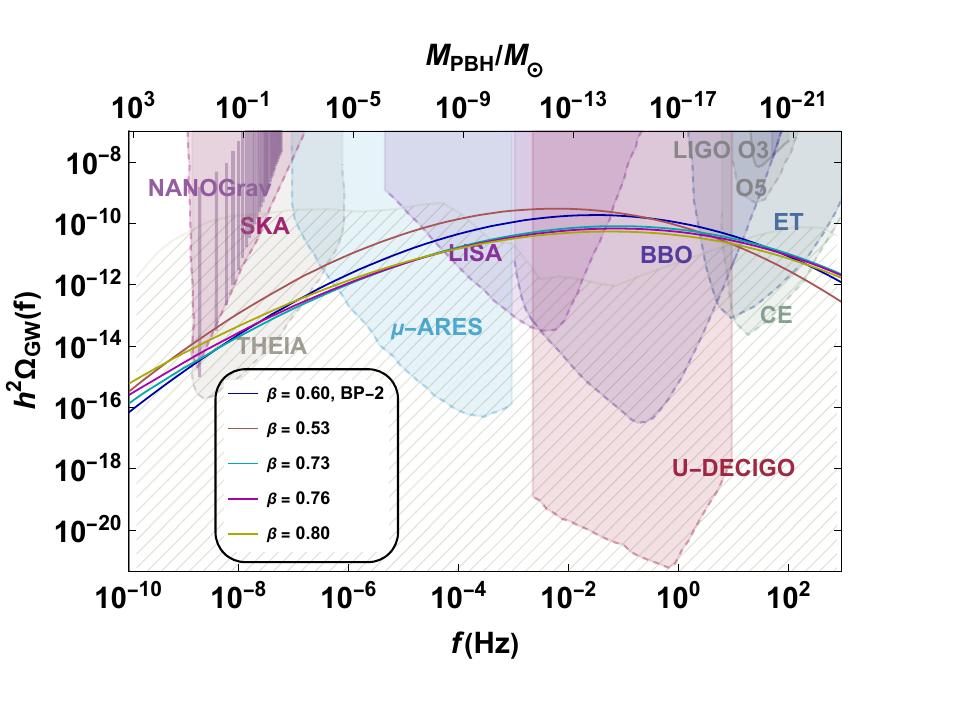}
    \quad
    \includegraphics[width=0.484\linewidth,height=5.7cm]{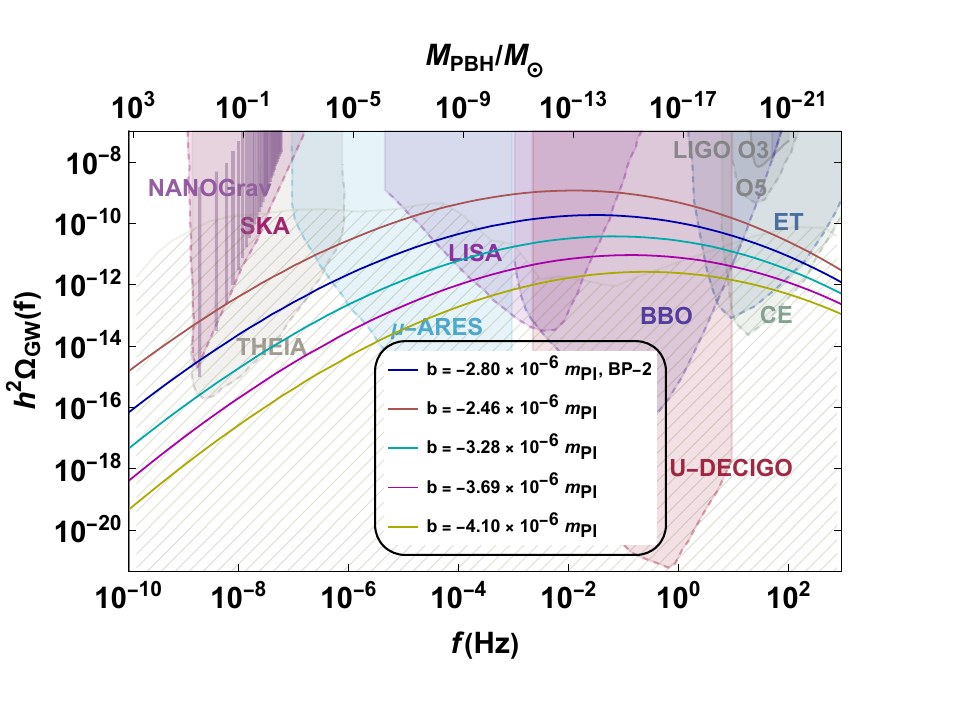}
\caption{\it The energy density of gravitational waves for \cref{Energdens} for the BPs given in \cref{parmsets}. The colored shaded regions indicate the sensitivity curves of present (solid boundaries) LIGO O3 \cite{KAGRA:2021kbb}, NANOGrav \cite{NANOGrav:2023gor} and future (dashed boundaries) LIGO O5, SKA \cite{Smits:2008cf}, THEIA \cite{Garcia-Bellido:2021zgu}, LISA \cite{Baker:2019nia}, $\mu$-ARES \cite{Sesana:2019vho}, BBO \cite{Corbin:2005ny}, U-DECIGO \cite{Yagi:2011wg, Kawamura:2020pcg}, CE \cite{Reitze:2019iox} and ET \cite{Punturo:2010zz} experiments. The hatched region shows the astrophysical background 
\cite{ghoshal2023traversing}.}
    \label{fig:Omegaf}
\end{figure}

The predicted $n_s$ and $r$ are shown in \cref{fig:rnsalpha}. The field evolution from the pivot scale to the end of inflation using \cref{bac} is shown in \cref{fig:phipsi}. The scalar power spectrum \cref{Powspec} with the variation of different model parameters is presented in \cref{fig:PS_k} along with the current and future experimental bounds as explained in \cref{realsiticframe}. The relevant PBH abundance is given in \cref{fig:f_pbh} along with the present and future experimental bounds (see \cref{sec:Modelcomp} for details). The SGWB formed by the PBH and the variation of different model parameters along with the experimental sensitivities (explained in \cref{sec:SIGW}) is given in \cref{fig:Omegaf}. It is important to note that the large value of the coupling $\beta$ predicted in \cref{sec:Modelcomp} is now taking a natural value in the $\alpha$-attractor scenario but the mass scale $M$ is larger than the reduced Planck scale and is less than the Planck mass, see \cref{parmsets}. 

\medskip

\section{Fine-tuning Estimate}
\label{sec:finetune}
The single-field inflationary models usually require a lot of fine-tuning of the parameters involved in the enhancement of the power spectrum at small scales \cite{Ballesteros:2017fsr}. However, in hybrid inflation due to the presence of another field, the amount of fine-tuning reduces significantly. The fine-tuning  parameterization in terms of a quantity $\Delta_x$ is given by
\begin{align}
\label{finetuneq}
    \Delta_x = \text{Max}\bigg|\dfrac{\partial \ln P_s^\text{Peak}}{\partial \ln x}\bigg|.
\end{align}
Here, $x$ is the underlying model parameter. The larger the $\Delta_x$ is, the larger the amount of required fine-tuning. Let us separately discuss each model we considered.

\section*{Model$-1$}
In this framework, $x\in\{M,\phi_c,m_1\}$ in \cref{finetuneq}. Evaluating numerically, fine-tuning estimates for toy model parameters \cref{potentialCl} are given in \cref{finetuneCl}. The maximum fine-tuning we obtain is $3$, which is almost six orders of magnitude smaller than single-field inflation \cite{Stamou:2021qdk} and two orders of magnitude smaller than standard hybrid inflation previously explored \cite{Spanos:2021hpk}.
\begin{table}[tbh!]
	  \fontsize{12pt}{13pt}
	\caption{ \it Fine-tuning (FT) estimate of model parameters for BP in \cref{parmsetsCl} with a peak of the spectrum around $5\times 10^{-3}$.  } 
	\centering 
	\begin{adjustbox}{max width=\columnwidth}
		\begin{tabular}{|M{1.5cm} |M{1cm} |M{1cm} |M{1cm} |M{1cm}|}
			\hline
   \bf{$\Delta_x$}& \bf{$\Delta_M$}&\bf{$\Delta_{\phi_c}$}&\bf{$\Delta_{m_1}$}
			\\ [0.6ex] 
			\hline\hline
			$\text{FT}$ &$3$  &$1$  & $-$\\
			\hline 
		\end{tabular}
	\end{adjustbox}
	\label{finetuneCl}
\end{table}
We find the fine-tuning estimate for $m_1$ is negligible.
\section*{Model$-2$}
In this framework, $x\in\{M,\beta,\gamma,\lambda_N\}$ in \cref{finetuneq}. Evaluating numerically, fine-tuning estimates for sneutrino theory parameters \cref{snpotencanon} are given in \cref{finetune}.  The maximum fine-tuning we obtain is $8$, which is almost five orders of magnitude smaller than single-field inflation \cite{Stamou:2021qdk} and one order of magnitude smaller than standard hybrid inflation \cite{Spanos:2021hpk}.
\begin{table}[tbh!]
	  \fontsize{12pt}{13pt}
	\caption{ \it Fine-tuning (FT) estimate of model parameters for BP$-1$ in \cref{parmsets} with a peak of the spectrum around $5\times 10^{-3}$.  } 
	\centering 
	\begin{adjustbox}{max width=\columnwidth}
		\begin{tabular}{|M{1.5cm} |M{1cm} |M{1cm} |M{1cm} |M{1cm}|M{1cm}|M{1cm}|M{1cm}|M{1cm}|M{1cm}|M{1cm}|M{1cm}|}
			\hline
   \bf{$\Delta_x$}& \bf{$\Delta_{M}$}&\bf{$\Delta_\beta$}&\bf{$\Delta_{\gamma}$}&\bf{$\Delta_{\lambda_N}$}
			\\ [0.6ex] 
			\hline\hline
			$\text{FT}$ &$8$  &$3$  & $1$ & $4$\\
			\hline 
		\end{tabular}
	\end{adjustbox}
	\label{finetune}
\end{table}

\medskip

\section{Discussion and Conclusion}
\label{sec:concl}

In summary, we presented two models of supersymmetric hybrid inflation (with sneutrino playing the role of the inflaton) and investigated the predictions of gravitational waves and primordial black holes generated during the waterfall transition which poses an added advantage of very mild fine-tuning unlike single field inflation \cite{Stamou:2021qdk}. We predict spectral index $n_s$ to be $0.966$ and the tensor-to-scalar ratio $r\simeq10^{-11}$ for the first analysis based on the assumptions of the toy model. For the $\alpha$-attractor case, ($n_s\simeq0.961-0.969$) and ($r\simeq0.0056-0.0068$).  These predictions are consistent with the current Planck data and within the reach of next-generation CMB experiments like LiteBIRD, etc and also satisfy PBH as dark matter with detectable GW signal. The inflaton is the lightest singlet sneutrino, so it dominates the reheating after inflation. Using $\langle\phi\rangle = \sqrt{{M'} M}$, we find 
its mass to be  $ 2 (\lambda_N)_{11} {M'}M/M_*$. 
It decays mainly via the extended MSSM 
Yukawa coupling into 
slepton and Higgs or a lepton and Higgsino with a decay width given 
by $\Gamma_{\Tilde{N}} = M_\mathrm{\Tilde{N}} {(Y^\dagger_\nu Y_\nu)_{11}}/{(4\pi)}$. 
The decay of the singlet sneutrino after inflation is responsible for reheating the visible universe to a 
temperature $T_{\mathrm{RH}} \approx (90/(228.75\pi^2 ))^{1/4} \sqrt{\Gamma_{\Tilde{N}} m^{}_{\mathrm{P}}}$. Following this, the typical reheat temperature for both the models is around $10^{7}-10^{8}$ GeV suitable for non-thermal leptogenesis as shown in \cref{table1,table2}.

Particularly we were able to show a novel correlation between the mass of PBH, the peak in the GW spectrum, and the right-handed (s)neutrino mass in both the models. The salient features of our analysis are:
\\
\\
{\it Model$-1$}:\\
\begin{itemize}
\item We presented a toy model of hybrid inflation by adding the linear term that helps to avoid PBH overproduction. We derived an analytical expression for the power spectrum that fits very well with the numerical results up to a certain range of the coefficient of the linear term (see \cref{fig:AnaExactPS_k_Fudge_factor}).
\item We achieved acceptable values for both spectral index ($n_s\simeq0.966$) and tensor-to-scalar ratio ($r\simeq 10^{-11}$) satisfying PBH as the entire dark matter of the universe and detectable GW signals.
    \item We compared the toy model with the sneutrino canonical hybrid inflation potential where the inflation is driven by the sneutrino. The BPs are given in \cref{parmsetsSNCL}. The model predicts a sub-Planckian mass parameter $M$, while the coupling between a gauge singlet and the waterfall field $\beta$ is $O(10^2)$.
    \item Second-order tensor perturbations propagating as GWs are predicted with amplitude $\Omega_{\rm GW}h^2$ $\sim 10^{-9}$ and peak frequency $f\sim 0.1$ Hz by LISA and $\Omega_{\rm GW}h^2 \sim 10^{-11}$ and peak frequency of $\sim 10$ Hz in ET (see \cref{fig:OmegafSNCl}). The production of PBH of mass around $ 10^{-13} M_\odot$ as the sole DM candidate in the universe is proposed. This novel candidate for DM is also a signature of the sneutrino (see \cref{fig:f_pbh}).
    \item The fine-tuning for this model is almost $O(1)$ or smaller (see \cref{finetuneCl}) which shows its supremacy over the single-field inflationary model.
    \end{itemize}
{\it Model$-2$}:\\
    \begin{itemize}
    \item We modify the sneutrino potential by introducing an $\alpha$-attractor transformation and see that the coupling $\beta$ takes a naturally small value but the mass parameter $M$ is larger (see \cref{parmsets}).
   \item We achieved acceptable values for both the spectral index ($n_s\simeq0.961-0.969$) and tensor-to-scalar ratio ($r\simeq 0.0056-0.0068$) satisfying PBH as dark matter and detectable GW signal.
    \item Second-order tensor perturbations propagating as GWs are predicted with amplitude $\Omega_{\rm GW}h^2$ $\sim 10^{-9}$ and peak frequency f $\sim 0.1$ Hz by LISA and $\Omega_{\rm GW}h^2 \sim 10^{-11}$ and peak frequency of $\sim 10$ Hz in ET in this model (see \cref{fig:Omegaf}). The production of PBH of mass around $ 10^{-13} M_\odot$ as the sole DM candidate in the universe is proposed. This novel candidate for DM is also a signature of the sneutrino (see \cref{fig:f_pbh}).
    \item The fine-tuning for this model is almost $O(1)$ (see \cref{finetune}) which is less than the hybrid inflation previously explored in the literature \cite{Spanos:2021hpk,Afzal:2024xci}.

\end{itemize}
Thus, we offer one potential way to test the origin of right-handed neutrino mass generation, which is currently inaccessible in colliders, using a GW detector.
Thus we find that if the cosmic inflation is driven by the sneutrino responsible generation of right-handed neutrino mass, one could see its imprints in gravitational lensing observations and PBH as the entire DM candidate and be tested in upcoming GW experiments.

As a future outlook, it could be interesting to study the impact of non-Gaussianities during the waterfall transition in the models studied as they impact the abundance of PBH formation rate, 
PBH clustering, and the amplitude of scalar-induced GW signals (see ref.~\cite{Domenech:2021ztg, Pi:2024jwt} for recent reviews). If some characteristic features of the GW spectral shapes encountered in this study are observed, one may look to target additional observations to distinguish between SUSY-mediated inflation and other scenarios. Particularly in sneutrino masses of TeV, these could be searched in experiments \cite{ParticleDataGroup:2022pth} at the particle physics laboratories. In this manner, we can complement GW searches with laboratory searches in the same BSM parameter space.
Gravitational wave astronomy with the planned global network of GW detectors aspires to achieve measurement precisions that are orders of magnitude better than the present-day detectors. This new era of GW detectors worldwide will make the dream of testing fundamental BSM mechanisms, e.g. for supersymmetry physics, or neutrino physics of the universe and inflationary cosmology, a reality in the near future.

\section*{Acknowledgements}

The authors thank Ahmad Moursy, Koushik Dutta, Konstantinos Dimopoulos, Rathul Raveendran, and Charalampos Tzerefos for the helpful comments. AA thanks Shahid Beheshti University Iran for the hospitality.

\appendix
\section{Effect of Non-linearities on PBH abundance}
\label{appen1}
\begin{figure}[tbh!]
    \centering
    \includegraphics[width=0.6\linewidth]{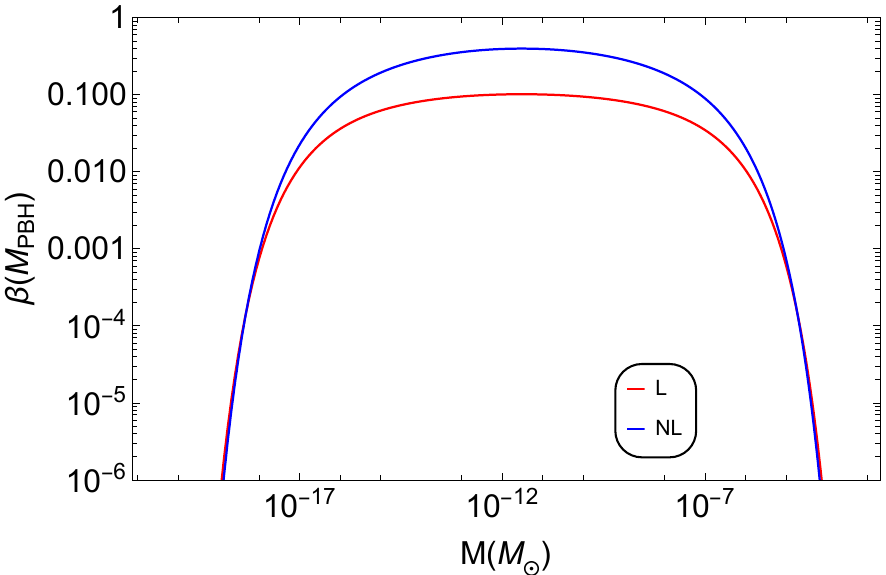}
\caption{\it Comparison of PBH abundance with (NL) and without (L) non-linearities.}
    \label{fig:betacompare}
\end{figure}

The non-linear relation between curvature perturbations and the density contrast field, the PBH fractional energy density can be written as \cite{Frosina:2023nxu}
\begin{align}
\label{betaNL}
\beta(M_\text{PBH})=\mathbb{K} \int_{\delta_L^\text{min}}^{\delta_L^\text{max}}\left(\delta_L-\dfrac{\delta_L^2}{4\,\Phi}-\delta_c\right)^{\gamma_c}\, P_G(\delta_L)\,d\delta_L.
\end{align}
Here, $\delta_L$ represents the linear Gaussian component of the density contrast field and
\begin{align}
 P_G(\delta_L)=\dfrac{1}{\sqrt{2\pi}\sigma(M_H)} e^{-\delta_L^2/2\,\sigma^2(M_H)}, \,\,\,\,\, M_H(T)=1.5\times10^5\left(\dfrac{g_\star(T)}{10.75}\right)^{-1/2}\left(\dfrac{T}{\text{MeV}}\right)^{-2}M_\odot
\end{align}
Following \cite{Frosina:2023nxu}, other constants are defined to be $\mathbb{K}=4.36, \Phi=2/3$ and $\gamma_c=0.38$. We compare the results without nonlinear effects \cref{beta} and with non-linear effects \cref{betaNL} in \cref{fig:betacompare} and show as an example, the same for a BP in \cref{parmsetsCl}. Since the difference between the two approaches is less than $O(1)$ we do not expect the results we present to change too much. Therefore, we view these as uncertainties in the calculation of the PBH abundance, just as the uncertainties in other parameters, like $\gamma_c$ which depends on the details of gravitational collapse \cite{Ando:2018qdb} and $\delta_c$ \cite{DeLuca:2023tun}.
\newpage

\bibliography{Bibliography}
\bibliographystyle{JHEP}

\end{document}